\documentclass{aa}
\usepackage[dvips]{graphicx}
\usepackage{psfrag}

\usepackage{natbib}

\def\lesssim{\mathrel{\hbox{\rlap{\hbox{\lower4pt\hbox{$\sim$}}}\hbox{$<$}}}}
\def\gtrsim{\mathrel{\hbox{\rlap{\hbox{\lower4pt\hbox{$\sim$}}}\hbox{$>$}}}}

\begin{document}

\title{Radio emission from cosmic ray air showers}

\subtitle{Monte Carlo simulations}

\author{T. Huege\inst{1} \and H. Falcke\inst{1,2,3}}

\offprints{T. Huege,\\email: thuege@mpifr-bonn.mpg.de}

\institute{Max-Planck-Institut f\"ur Radioastronomie,
           Auf dem H\"ugel 69, 53121 Bonn, Germany
         \and
           Radio Observatory, ASTRON, Dwingeloo, P.O. Box 2, 7990 AA Dwingeloo, The Netherlands
         \and
           Adjunct Professor, Dept. of Astronomy, University of Nijmegen, P.O. Box 9010, 6500 GL Nijmegen, The Netherlands
}

\date{Received August 20, 2004; accepted October 10, 2004}

\abstract{ We present time-domain Monte Carlo simulations of radio emission from cosmic ray air showers in the scheme of coherent geosynchrotron radiation. Our model takes into account the important air shower characteristics such as the lateral and longitudinal particle distributions, the particle track length and energy distributions, a realistic magnetic field geometry and the shower evolution as a whole. The Monte Carlo approach allows us to retain the full polarisation information and to carry out the calculations without the need for any far-field approximations. We demonstrate the strategies developed to tackle the computational effort associated with the simulation of a huge number of particles for a great number of observer bins and illustrate the robustness and accuracy of these techniques. We predict the emission pattern, the radial and the spectral dependence of the radiation from a prototypical $10^{17}$~eV vertical air shower and find good agreement with our analytical results \citep{HuegeFalcke2003a} and the available historical data. Track-length effects in combination with magnetic field effects surprisingly wash out any significant asymmetry in the total field strength emission pattern in spite of the magnetic field geometry. While statistics of total field strengths alone can therefore not prove the geomagnetic origin, the predicted high degree of polarisation in the direction perpendicular to the shower and magnetic field axes allows a direct test of the geomagnetic emission mechanism with polarisation-sensitive experiments such as LOPES. Our code provides a robust, yet flexible basis for detailed studies of the dependence of the radio emission on specific shower parameters and for the inclusion of additional radiation mechanism in the future.
}

\maketitle


\section{Introduction}

Extensive air showers initiated by high-energy cosmic rays have been known for almost 40 years to produce strongly pulsed radio emission at frequencies from a few to a few hundred MHz (\citealt{JelleyFruinPorter1965}; for an excellent review see \citealt{Allan1971}). The quality of the experimental data as well as the theoretical understanding of the emission mechanism, however, stagnated on a rather basic level when the research in this field almost ceased completely in the late 1970ies.

Lately, the advent of new fully digital radio interferometers such as LOFAR\footnote{http://www.lofar.org} sparked renewed interest in the topic. The radio technique potentially offers a cost-effective additional method for the study of extensive air showers which would very much complement the classical particle detector techniques at moderate cost \citep{FalckeGorham2003}. Out of this interest, the LOPES project was initiated as a joint effort to firmly establish the feasibility of radio measurements of extensive air showers with a LOFAR prototype station \citep{HornefferFalckeHaungs2003} operating in conjunction with the KASCADE experiment \citep{AntoniApelBadea2003} as well as to gain a solid theoretical understanding of the emission mechanisms.

In \citet{HuegeFalcke2003a} we presented calculations of radio emission from extensive air showers within the scheme of ``coherent geosynchrotron radiation" first proposed by \citet{FalckeGorham2003}. These calculations were based on an analytic approach and were specifically aimed at gaining a solid understanding of the coherence effects that shape the radiation emitted by an air shower. Building on this foundation, we now continue to develop and enhance our model with elaborate Monte Carlo (MC) simulations. The MC technique allows us to infer the emission characteristics with much higher precision by taking into account more realistic and complex shower properties and applying fewer approximations than in the analytic calculations. At the same time, it provides an independent means to verify our previous calculations due to the totally different technique employed. The approach we take is similar to the MC simulations done by \citet{SuprunGorhamRosner2003}, yet our simulation is developed to a much higher level of complexity.

The layout of this paper is as follows: In sections 2 and 3 we motivate and explain the application of the MC technique and provide details about its implementation. In section 4, we explain the ``intelligent'' concepts that we have explored to make simulations with a high number of particles and observer bins feasible on standard personal computer hardware. After a short description of the raw output of our program and the associated data reduction in section 5, we demonstrate the robustness and consistency of the implemented algorithms in detail in section 6. Similar to our earlier theoretical calculations, we then first concentrate on the emission from a single ``slice'' of particles in the air shower and compare the results with the analytical results in section 7. Next, we perform the integration over the shower evolution as a whole and present the results in comparison with our analytical work and historic data in section 8 before concluding with a discussion of the results and our conclusions in sections 9 and 10, respectively.


\section{The Monte Carlo approach}

After having established the general dependencies of the radio emission on a number of air shower properties in \citet{HuegeFalcke2003a}, a continuation of the calculations with MC techniques offers a number of advantages.

\subsection{Motivation and objectives}

First, MC techniques allow an independent verification of the analytic calculations by adopting the exact same shower properties (e.g., distributions of particles in space and energy, choice of magnetic field and shower geometry), but doing the calculations in a completely different way. In particular, the MC calculations are carried out by summing up the individual particles' pulses in the time-domain, whereas the analytic calculations were done in the frequency domain.

Second, it is relatively easy to include highly complex (and thus realistic) shower characteristics in the MC simulations, thereby increasing the model precision significantly over the previous results. Additionally, it is not necessary to make approximations such as adopting a far-field limit, which reduces the accuracy of the analytic calculations. 

Overall, MC simulations therefore constitute the logical next step in the development of our model.

\subsection{General approach} \label{sec:mcapproach}

The general idea of a MC simulation of radio emission from cosmic ray air showers is simple:
\begin{enumerate}
\item{model the radiation emitted by an individual particle as precisely as possible,}
\item{distribute particles randomly in a simulated shower according to the desired shower characteristics (e.g., spatial and energy distributions),}
\item{superpose the radiation received from the shower particles at the given observing positions, taking into account retardation effects.}
\end{enumerate}
In fact, the ``microphysics" of an individual particle's geosynchrotron emission (step 1) can be described analytically without the need of any approximations as long as one does not simulate particle interactions explicitly but only takes them into account via statistically distributed track lengths. The strategy for the implementation of steps 2 and 3 then is as follows:
\begin{itemize}
\item{generate shower particles according to the desired distributions,}
\item{for each particle:}
\begin{itemize}
\item{for each observing position (ground-bin):}
\begin{itemize}
\item{establish an adequate sampling of the particle trajectory,}
\item{calculate and retard the emission contributions emanating from the sampled points on the trajectory, building up the electric field time-series that the observer sees,}
\item{incorporate the contributions into the ground-bin's pre-existing time-series data.}
\end{itemize}
\end{itemize}
\end{itemize}
This strategy is fairly simple. What makes the problem difficult is the huge computational effort of a simulation with a large number of particles and ground-bins, as was already discussed by \citet{DovaFanchiottiGarciaCanal1999} who considered a similar approach. A number of intelligent concepts has to be applied to actually develop a working simulation out of this simple ``recipe''. We discuss these concepts in depth in section \ref{sec:intconcepts}.


\section{Implementation details} \label{sec:impdetails}

In this section we describe a number of relevant implementation details of our MC code.

\subsection{Technical information}

The program has been developed under Linux using gcc 2.9.5, gcc 3.3.1, and the Intel C++ compiler version 8.0.5 which produces much faster code. No third-party libraries were used in order to maximise portability and minimise dependency on external factors. The program source-code will be made available at a later time.

\subsection{Particle creation and propagation}

The particles in the shower are created with random properties distributed according to analytic parametrisations. If not explicitly stated otherwise, the parametrisations are chosen exactly as in \citet{HuegeFalcke2003a}, to which we refer the reader for detailed information. While these parametrisations are admittedly crude and do not take into account some air shower properties such as a realistic particle pitch-angle distribution, this approach allows a direct comparison of the MC results with our analytic calculations. A more realistic modelling of the air shower will be achieved once our code is interfaced with the air shower simulation code CORSIKA. The particles are created with the following properties chosen randomly:
\begin{itemize}
\item{the shower age at which a particle is created (longitudinal development according to \citet{Greisen1960} function), which directly yields}
\begin{itemize}
\item{the position along the shower axis}
\item{the creation time}
\end{itemize}
\item{the lateral shift from the shower axis (NKG-distribution dating back to \citet{KamataNishimura1958} and \citet{Greisen1960})}
\item{the longitudinal shift along the shower axis as function of the lateral shift (asymmetrical $\Gamma$-PDF)}
\item{the azimuth angle for the lateral shift (isotropic)}
\item{the particle gamma factor (broken power-law distribution or fixed $\gamma \equiv 60$)}
\item{the particle track length (exponential probability distribution or fixed $\lambda \equiv 40$~g~cm$^{-2}$)}
\end{itemize}
To take into account the pair-wise creation of particles, one electron and one positron are always generated with the same properties. At the moment, no random spread in the particle pitch-angle is introduced, i.e., the initial particle momenta radially point away from a spherical surface with 2,300~m radius, exactly as in the analytical calculations, motivated by the data from \citet{AgnettaAmbrosioAramo1997}. The fact that the initial velocity direction is shared by both the electron and positron introduces only minor error as the transverse momentum arising from the pair production is minimal --- a fact that is intuitively illustrated by the still very dense core of the lateral distribution function even after many generations of particle creation.

As each of the particles is created at a specific position at a given time with a given initial velocity, the trajectory $\vec{r}(t)$ resulting from the deflection in the geomagnetic field is a well-defined helix which can easily be described analytically by equations (\ref{eqn:trajecr}), (\ref{eqn:trajecrabs}) and (\ref{eqn:trajecr0}) as derived in the appendix.

\subsection{Calculating and collecting contributions}

Once the trajectory (and its time-derivatives) are known analytically, the radiation an observer at position $\vec{x}$ receives at time $t$ can be calculated (cf. \citealt{Jackson1975} equation 14.14) immediately by
\begin{eqnarray} \label{eqn:radiate}
\vec{E}(\vec{x},t) &=& e \left[\frac{\vec{n}-\vec{\beta}}{\gamma^{2}(1-\vec{\beta} \cdot \vec{n})^{3} R^2}\right]_{\mathrm{ret}} \nonumber \\
&+& \frac{e}{c} \left[ \frac{\vec{n} \times \{(\vec{n}-\vec{\beta})\times \dot{\vec{\beta}}\}}{(1-\vec{\beta}\cdot\vec{n})^{3}R}\right]_{\mathrm{ret}},
\end{eqnarray}
where $e$ denotes the particle charge, $\vec{\beta}(t) = \vec{v}(t)/c$ is directly given by the particle velocity, $\vec{R}(t)$ refers to the vector between particle and observer position, $R(t)=\left|\vec{R}(t)\right|$ and $\vec{n}(t)=\vec{R}(t)/R(t)$ denotes the line-of-sight direction between particle and observer.

The index ``ret'' points out that the quantities in the brackets have to be evaluated at the {\em{retarded}} time
\begin{equation} \label{eqn:retardation}
t_{\mathrm{ret}}= t-R(t_{\mathrm{ret}})/c
\end{equation}
rather than at the time $t$ in order to accommodate the finite light-travel time. This ``recursive'' retardation relation imposes significant problems for an analytical calculation in the time-domain. In case of a MC simulation, on the other hand, it is absolutely straight forward to take the retardation into account by simply delaying the emitted signal appropriately before collecting it in the ground-bins.

The first term in equation (\ref{eqn:radiate}) constitutes the ``static'' term that falls off with $R^{-2}$ in field strength. It is usually neglected, and was not taken into account in \citet{HuegeFalcke2003a} either. While its contributions are indeed negligible, we still include it in our MC simulation as to not make any unnecessary approximations.

The second term is the usual ``radiation'' term which drops as $R^{-1}$ in field strength and therefore dominates very quickly over the static term as one goes to higher distances. In analytic approaches, it is usually necessary to apply approximations such as the Fraunhofer-approximation for the far-field limit to this term, which naturally limits the precision of the results. Again, for a MC simulation, it is not necessary to make any approximation for the radiation formula.

Beaming effects are naturally taken into account in this formula through the $(1-\vec{\beta}\cdot\vec{n})^{3}$ terms in the denominator that lead to very high field strengths for particle velocities close to $c$ and small angles to the line-of-sight. As soon as one takes into account the refractive index of a medium rather than vacuum, the denominator actually becomes zero at the \v{C}erenkov angle. The arising singularity, in combination with the modified retardation relation, then leads to \v{C}erenkov radiation. The analysis of these effects, however, is beyond the scope of this work and will be carried out in a later paper.

For a given particle, the trajectory is then sampled in a sufficiently high number of points, and the retarded contribution emanating from each of these points is calculated for each of the observing ground-bins with the full precision of equation (\ref{eqn:radiate}). Thus, the retarded electric field time-series produced by the particle is inferred for each of the ground-bins and can then be incorporated into their pre-existing time-series data. The details and subtleties of this procedure are explained in sections \ref{sec:smartsampling} and \ref{sec:griddingstrategy}.

\subsection{Atmosphere model}

We use the US standard atmosphere of 1977 as implemented in CORSIKA \citep{Ulrich1997}. It splits the atmosphere in a number of layers, in each of which the path depth in g~cm$^{-2}$ (as counted vertically from the outer edge of the atmosphere) is parametrised by an exponential dependence on height
\begin{equation}
X(h) = a_{i} + b_{i}\ \exp \left( -\frac{h}{c_{i}} \right).
\end{equation}
The layer boundaries and corresponding parameters $a_{i}$, $b_{i}$ and $c_{i}$ are given in table \ref{tab:atmospherelayers}. The Moli\`{e}re radius $r_{\mathrm{M}}$ is parametrised as \citep{DovaEpeleMariazzi2003}
\begin{equation}
r_{\mathrm{M}}(X) = \frac{9.6}{(X - a_{i})} c_{i}.
\end{equation}

\begin{table}
  \begin{tabular}{ccccc}
     \hline
     Layer & Height [km] & $a_{i}$~[g~cm$^{-2}$] & $b_{i}$~[g~cm$^{-2}$] & $c_{i}$~[cm] \\
     \hline
     1 & 0 -- 4 & -186.56 & 1222.66 & 994186.38 \\
     2 & 4 -- 10 & -94.92 & 1144.91 & 878153.55 \\
     3 & 10 -- 40 & 0.61 & 1305.59 & 636143.04 \\
     4 & 40 -- 100 & 0.00 & 540.18 & 772170.16 \\
     \hline
  \end{tabular}
  \caption{
  \label{tab:atmospherelayers}
  Parameters for the parametrisation of the atmospheric layers.}
\end{table}

\subsection{Random number generation}

An important centre-piece of any MC simulation is the random number generator at its heart. We revert to the well-known ``Mersenne-Twister'' random-number generator of \citet{MatsumotoNishimura1998} in the C++ implementation by \citet{Fog2003}. If possible, the generators for non-uniform probability distributions were implemented using analytic inversions. If not, we reverted to rejection methods.


\section{Intelligent concepts} \label{sec:intconcepts}

As mentioned earlier, a ``brute-force'' approach as sketched in the ``recipe'' given in section \ref{sec:mcapproach} involves such a high computational effort (both regarding CPU time and memory) that it simply is not feasible on standard PCs. One therefore has to employ a number of intelligent concepts to minimise the computing effort. The main ideas that we have explored to reach this goal are described in the following subsections.

\subsection{Cutting off $\gamma^{-1}$-cones} \label{sec:cutoffgamma}

\begin{figure}
   \begin{center}
   \includegraphics[width=7.0cm]{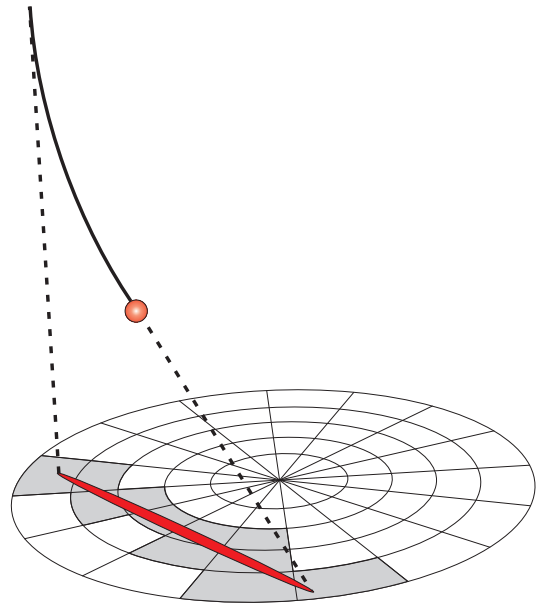}
   \end{center}
   \caption{
   \label{fig:groundtrace}
   Cutting off $\gamma^{-1}$-cones: only bins in the ``ground-trace'' of a particle's trajectory are selected for evaluation.
   }
   \end{figure}

The radiation pattern emitted by an individual highly relativistic particle is heavily beamed in the forward direction (see, e.g., \citealt{Jackson1975}). Most of the emission is radiated into a cone with opening angle of order $\sim \gamma^{-1}$. This directly leads to a very simple, but effective idea of how to cut down on computing time:

For an individual particle flying on its given trajectory, the $\gamma^{-1}$ emission cone sweeps over a relatively small region on the ground. It is only in this ground-region of a few times $\gamma^{-1}$ ``width'', which we call the ``ground-trace'' of the particle trajectory, that receives considerable contributions of radiation from that particle. Thus, we can select only the ground-bins inside the ground-trace of this specific particle for evaluation as sketched in figure \ref{fig:groundtrace}, hugely cutting down on computing time.

It turns out, however, that the discrete cutting off after a certain angular distance introduces errors in the calculation, which are --- although only at a percent-level --- significant when one is interested in the emission strength at distances a few hundred metres from the shower core. We will discuss the details in section \ref{sec:evalgammacones}.

\subsection{Smart trajectory-sampling} \label{sec:smartsampling}

In general, there are two approaches of how to calculate the time-dependence of the emission that a specific ground-bin receives from a given particle:

The first approach is to take the trajectory of a given particle, sample it in a sufficient number of points and calculate the corresponding contributions for the given ground-bin using (\ref{eqn:radiate}). If the particle trajectory is sampled in equidistant time-steps, the retardation effects involved lead to a {\em{non-equidistantly sampled}} time-series in the ground-bin. The heavily-peaked pulse shape is, however, {\em{automatically}} sampled with high precision in this approach.

As the time-series data collected by the ground-bin have to be gridded eventually, it would be easier if one could take another approach: taking the ground-bin's pre-defined equidistant time-grid and sampling the particle trajectory in the points corresponding to this grid. This, however, is not easily possible because of the ``recursive'' retardation relation (\ref{eqn:retardation}). One would have to do an iterative search for the corresponding points on the particle trajectory, which would be at least as time-consuming as interpolating or binning a non-gridded time-series derived with the first approach to the ground-bin's time-grid. At the same time, this approach bears the risk of missing highly peaked contributions in case of a too widely spaced time-grid.

\begin{figure}
   \psfrag{t0ns}[c][t]{$t$~[ns]}
   \psfrag{E0muVpm}[c][t]{$\left|\vec{E}(t)\right|$~[$\mu$V m$^{-1}$]}
   \includegraphics[height=8.6cm,angle=270]{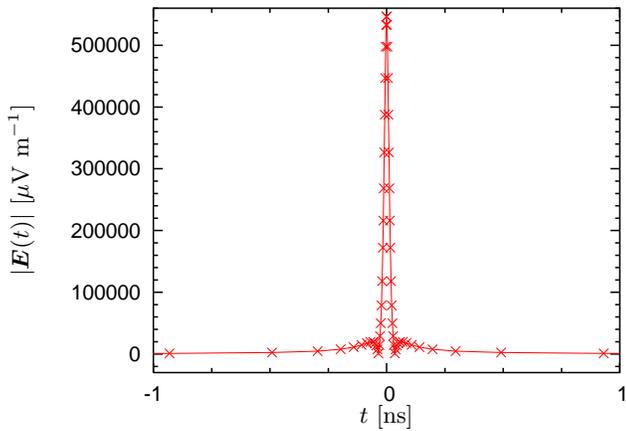}
   \caption{
   \label{fig:smartsampling}
   Smart trajectory sampling: the particle trajectory is sampled densely in the peak-region and only sparsely further outside.
   }
   \end{figure}

The first approach, therefore, is the better choice for our calculations. It is, however, possible to improve on the case of equidistant sampling of the particle trajectory by taking advantage of the strong beaming of the emission once again. The peaks in the time-series result when the denominator $(1-\vec{\beta}\cdot\vec{n})^{3}$ in (\ref{eqn:radiate}) gets small as the angle between $\vec{\beta}$ and $\vec{n}$ gets small. It therefore makes sense to densely sample the region of small angle to the line-of-sight, and increase the distance between the sampled points (up to a maximum value) as the angle gets larger. This minimises the number of points used while it guarantees high-precision sampling of the pulse shape as can be seen in figure \ref{fig:smartsampling}.

\subsection{Intelligent gridding strategy} \label{sec:griddingstrategy}

The individual particles' pulses are very short, of order 10$^{-11}$~s for $\gamma=60$ particles. Usually, however, one is only interested in the result on scales of several nanoseconds as defined by the interesting frequency range of tens to hundreds of MHz. On the other hand, it would be useful for diagnostic and verification purposes to have a means to record the events in the very high time-resolutions of the individual pulses. We have therefore implemented both possibilities into the simulation program. The different demands are met by the use of two very different gridding strategies.

The more efficient and thus favoured strategy for low time resolutions is the use of a ``simple grid''. It is based on an equidistant grid of user-defined resolution (typically $\sim 1$~ns). The contributions from a particle's time-series data are then binned and thus time-averaged onto the grid. The gridding is dynamic in the sense that data points are inserted automatically when needed. This saves memory in comparison with an ordinary equidistant grid. Still, this strategy does not scale well to high time-resolutions.

\begin{figure}
   \begin{center}
   \includegraphics[width=7.0cm]{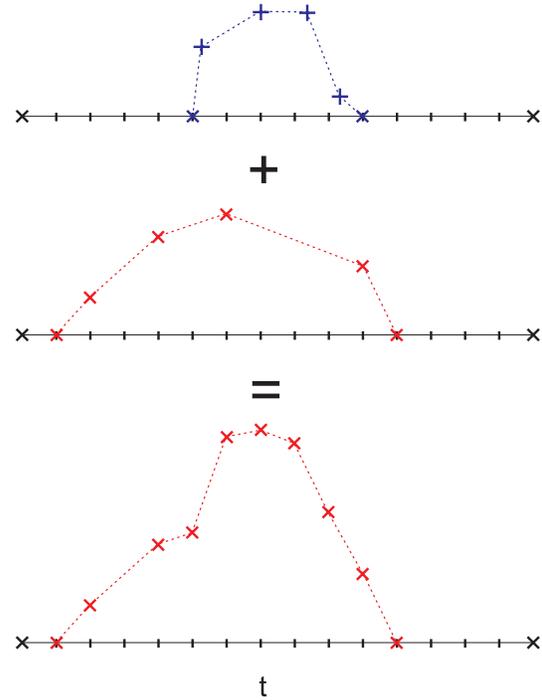}
   \end{center}
   \caption{
   \label{fig:economicgridding}
   The economic gridding mechanism: when new contributions are registered onto existing contributions, points are inserted and interpolated only as needed.
   }
   \end{figure}

To counter the high memory demands for high time resolutions, we have implemented an ``economic grid'' which describes the pulse-shapes using only a minimum number of points. It is based on an underlying equidistant time-grid which limits the time-resolution to a pre-defined maximum value. The time-series of an individual particle is interpolated to the underlying grid positions and then incorporated into the pre-existing time-series, correctly interpolating any contributions that were already present. Points are, however, only inserted in the grid where necessary. See Fig.\ \ref{fig:economicgridding} for an illustration of the algorithm.

As another major advantage, the availability of these two very different gridding strategies allows an independent cross-check of their implementation.

\subsection{Sequentialised and parallelised calculation}

The program is designed such that calculations can easily be sequentialised or parallelised to cut down on memory requirements or take advantage of multiple computers, respectively.

Memory usage increases with the number of ground-bins to be calculated. To facilitate the use of low-memory machines, the calculation of ground-bins can be sequentialised: the emission from the complete shower is calculated for only a subgroup of ground-bins, and only after the calculation has finished and the results have been written to disk, the next group of ground-bins is evaluated. This efficiently decreases the memory-usage as compared to concurrent evaluation of all ground-bins. The overhead introduced due to the necessary multiple creation of particle distributions is negligible for most combinations of parameters. Manually specifying the random seed value for the random number generator guarantees identical particle distributions in the different calculation segments. Allowing for different seed values in the calculations, on the other hand, provides a consistency check based on the underlying symmetries in the emission pattern as the different ground-bins are calculated based on independent sets of random numbers.

Similarly, different subgroups of the desired ground-area can be calculated in parallel on different computers,  yielding up to a linear decrease of net computation time.

\subsection{Automatic ground-bin inactivation}

As the simple discrete cutting off of regions outside the $\sim \gamma^{-1}$ ground-trace region described in section \ref{sec:cutoffgamma} introduces errors that are too big if one is interested in regions of a few hundred metres distance to the shower core, we developed a more sophisticated means of cutting down on computation time.

Since most of the particles are distributed in the innermost centre-region of the shower, the ground-bins close to the centre-region receive a high number of strong contributions, whereas the far-away regions only receive a smaller number of not-so-strong contributions. While the centre-regions might already have reached sufficient precision after a certain number of particles, the outer regions might still not have reached the desired statistical precision, affording a calculation with an even higher number of particles.

The computing time can be distributed in a much more efficient way by an on-the-fly inactivation of ground-bins that have reached the desired precision. To accomplish this, the program evaluates the shower emission in steps of (typically) 10,000 particles at a time. After each of these steps it compares the (smoothed) time-series derived for a specific ground-bin up to that stage with the results for that ground-bin at the previous step. Once the relative changes fall under a pre-defined limit for a user-defined number of steps in a row, the corresponding ground-bin is marked as ``inactive'' and is not evaluated any further. The computing time is thus effectively redistributed to the outer bins.

The fact that the inactivation sequence should propagate from the inner to the outer regions is used as a consistency check for the procedure. Another advantage is that the user can directly specify a desired precision rather than having to estimate the number of particles needed to reach adequate results.


\section{Data output and reduction}

To facilitate the understanding of the following discussions, we give a short overview of the raw data that the simulations produce and the kind of data ``reduction'' that we apply to visualise the results.

\subsection{Raw data}

\begin{figure}
   \psfrag{t0ns}[c][t]{$t$~[ns]}
   \psfrag{E0muVpm}[c][t]{$\left|\vec{E}(t)\right|$~[$\mu$V m$^{-1}$]}
   \includegraphics[height=8.6cm,angle=270]{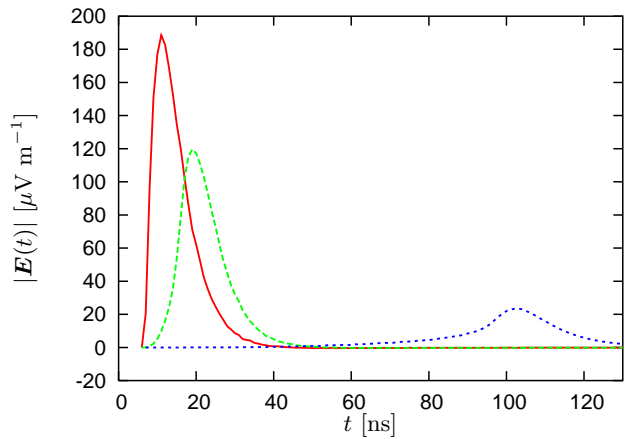}
   \caption{
   \label{fig:rawpulsesnorthpar03}
   Time-dependence of the raw pulses originating from the shower maximum as observed by an observer at (from left to right) 20~m, 140~m and 460~m to the north from the shower centre.
   }
   \end{figure}

The Monte Carlo code tracks the individual particles and calculates the associated (vectorial) electromagnetic pulses a specific observer (i.e., a specific ground-bin) receives. The individual pulses are extremely short, of order a few $10^{-11}$~s (cf.\ Fig.\ \ref{fig:smartsampling}). Superposition of all the individual pulses yields the raw output of the program: one data file per ground-bin stating the time-dependence of the north-south, east-west and vertical polarisation components of the electric field. Fig.\ \ref{fig:rawpulsesnorthpar03} shows the total field strength of the raw pulses at different distances to the north from the centre of a shower slice consisting of $10^{8}$ particles in the shower maximum. Due to the particle's spread in position and time, the pulses are considerably broader than the individual pulses, of order tens of nanoseconds.
%
%
A helper application is then used to process these raw data and reduce them to the desired physical quantities.

\subsection{Spectral filtering}

\begin{figure}
   \psfrag{Eomega0muVpmpMHz}[c][t]{$\left|\vec{E}(\vec{R},\omega)\right|$~[$\mu$V~m$^{-1}$~MHz$^{-1}$]}   
   \psfrag{nu0MHz}[c][t]{$\nu$~[MHz]}
   \includegraphics[height=8.6cm,angle=270]{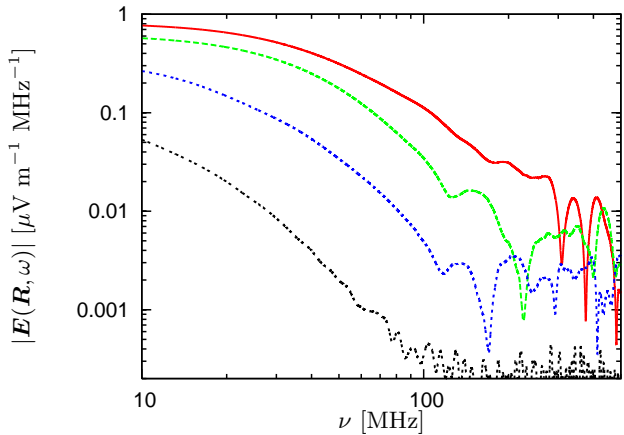}
   \caption{
   \label{fig:spectranorthpar03}
   Spectra of pulses originating from the shower maximum for observers at (from top to bottom) 20~m, 140~m, 340~m and 740~m distance to the north of the shower centre.
   }
   \end{figure}

In a first step, the time-series data of the electric field vector are Fourier-transformed, yielding the associated spectra depicted in Fig.\ \ref{fig:spectranorthpar03}. Due to coherence losses caused by interference effects, the spectra fall off steeply towards high frequencies. At a certain frequency, dependent on the distance from the shower centre, the field strength reaches a first interference minimum followed by a rapid series of alternating maxima and minima in the incoherent regime. Insufficient sampling of these extrema yields the unphysically seeming features seen here at high frequencies. In the emission from a real air shower such features are unlikely to exist as the inhomogeneities present in an air shower, but not taken into account in the analytic parametrisations, destroy the pronounced extrema. Calculation of the emission in this region therefore requires a more detailed air shower model, e.g. by interfacing of our code to CORSIKA.

\begin{figure}
   \psfrag{Eew0muVpm}[c][t]{$E_{\mathrm{EW}}(t)$~[$\mu$V m$^{-1}$]}
   \includegraphics[height=8.6cm,angle=270]{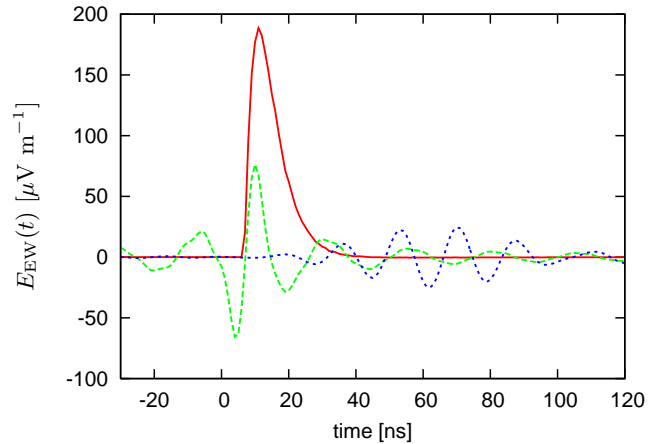}
   \caption{
   \label{fig:rawsmoothpulsespar03}
   Comparison of the east-west component of a raw pulse (solid), a pulse smoothed with a 40--160~MHz idealised rectangle filter (long dashed) and a 42.5--77.5~MHz filter as used in LOPES (short dashed) for emission from the shower maximum.
   }
   \end{figure}

Any concrete experiment will have a finite frequency bandwidth. Thus, we filter the spectra to infer the actual pulses that the experiment will register. As is obvious from the spectra, most of the power resides at the low frequencies. For this reason the amplitude drops significantly when filtering frequencies below, e.g., 40~MHz as seen in Fig.\ \ref{fig:rawsmoothpulsespar03}. 

Here we have used the same idealised 40--160~MHz filter that we applied in the theoretical calculations of \citet{HuegeFalcke2003a} as well as the actual 42.5--77.5~MHz filter that is used in LOPES. The acausality of the pulse filtered with the idealised rectangle filter illustrates that such a filter is an unphysical concept. A physical filter does not show acausal behaviour despite its steep edges as it delays the signal appropriately. This is well visible for the LOPES filter.

\subsection{Further data processing}

To analyse the radial dependence of the emission strength, we then determine the maximum amplitude of the filtered pulse in each ground-bin. The dependence of this ``filtered pulse maximum amplitude'' can then be visualised in a number of ways, e.g., as surface plots, contour plots or cuts in specific directions (see the following sections for examples). The absolute time associated to the maximum pulse amplitude additionally yields information about the curvature of the radio wave front.

One subtlety involved with this procedure is the noise levels present at high radial distances. As can be seen in Fig.\ \ref{fig:rawpulsesnorthpar03}, the pulses get significantly broader as one goes to higher distances. (This effect gets even stronger for fully integrated showers.) At distances of several hundred metres, the pulses are so broad that a filter clipping frequencies below $\sim$40~MHz actually resolves the pulses out. Consequently, the pulse amplitude drops to very low values comparable to those introduced by the higher-frequency numerical noise associated with the very short individual particle pulses. Taking the maximum amplitude as a measure for the emission strength then might no longer constitute a useful procedure. Calculating the time-integral of the field strength or the received power would be a better approach in these cases.

For the diagnostics of the employed algorithms in section \ref{sec:consistencychecks} and the analysis of track length effects in section \ref{sec:edgeeffects} we therefore adopt 800~m as a cutoff distance. For the calculation of emission strengths from a shower slice and an integrated shower, we limit the plots to distances of $\sim$550~m and $\sim$400~m, respectively, to insure that the filtered pulse amplitudes give adequate results which are directly related to the experimentally relevant signal-to-noise levels.


\section{Consistency checks} \label{sec:consistencychecks}

We have studied the output of our MC code very carefully to make sure that the calculations are correct. In particular, we have made the following consistency checks:

\subsection{Individual particle pulses} \label{sec:individualpulses}

\begin{figure}
   \psfrag{t0ns}[c][t]{$t$~[ns]}
   \psfrag{E0muVpm}[c][t]{$\left|\vec{E}(t)\right|$~[$\mu$V m$^{-1}$]}
   \includegraphics[height=8.6cm,angle=270]{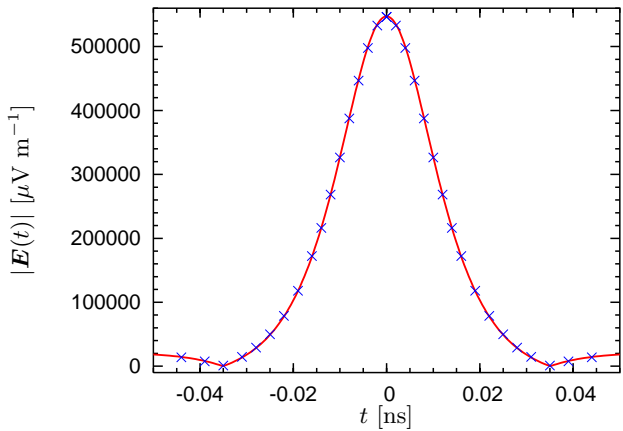}
   \caption{
   \label{fig:mcanalyticpulse}
   Total field strength of a pulse originating from a point source of $10^8$ $\gamma=60$ particles in 4~km height, observed in the shower centre. Solid: analytic calculation, points: calculated with MC code.
   }
   \end{figure}

Figure \ref{fig:mcanalyticpulse} shows the comparison between an analytical calculation and our MC code for a pulse created by a point-source consisting of $10^8$ particles with $\gamma \equiv 60$ at 4~km height, comparable to a pulse that would originate from the maximum of a vertical $10^{17}$~eV air shower concentrated into a point, as seen from an observer in the shower centre. The magnetic field is adopted as horizontal with a strength of 0.3~Gauss. The good agreement between the MC and analytic results demonstrates that the calculation of the particle trajectories and emission contributions is implemented correctly.

\begin{figure}
   \psfrag{t0ns}[c][t]{$t$~[ns]}
   \psfrag{E0muVpm}[c][t]{$E_{\mathrm{EW}}(t)$~[$\mu$V m$^{-1}$]}
   \includegraphics[height=8.6cm,angle=270]{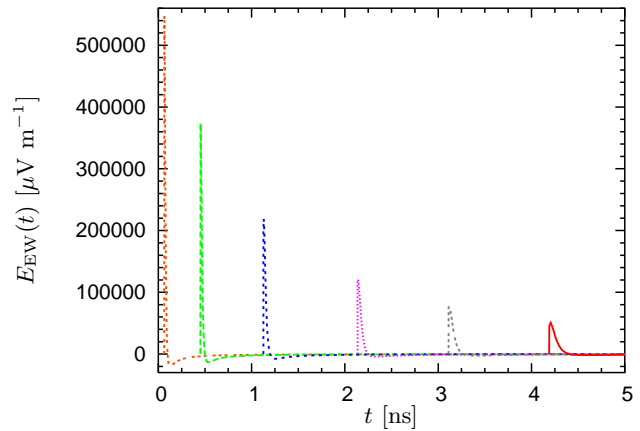}
   \caption{
   \label{fig:mcpulsesN}
   East-west polarisation component of a point-source with $\gamma=60$ particles at 4~km height at increasing distance {\em{to the north}} from the shower axis. From left to right: 5~m, 305~m, 505~m, 705~m, 855~m and 995~m.
   }
   \end{figure}

\begin{figure}
   \psfrag{t0ns}[c][t]{$t$~[ns]}
   \psfrag{E0muVpm}[c][t]{$E_{\mathrm{EW}}(t)$~[$\mu$V m$^{-1}$]}
   \includegraphics[height=8.6cm,angle=270]{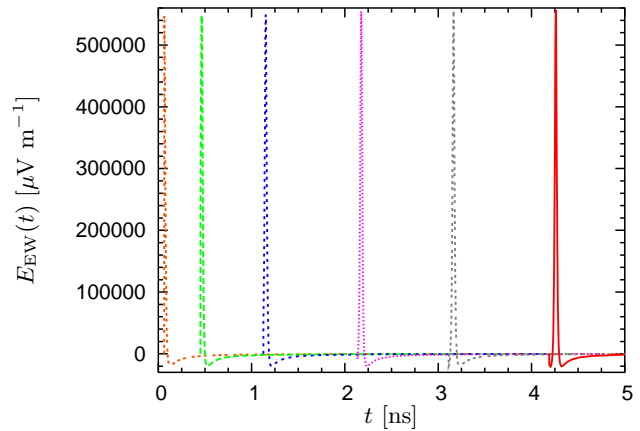}
   \caption{
   \label{fig:mcpulsesE}
   Same as figure \protect\ref{fig:mcpulsesN} at increasing distance {\em{to the east}} from the shower axis.
   }
   \end{figure}

Figures \ref{fig:mcpulsesN} and \ref{fig:mcpulsesE} demonstrate how the pulses from the same point-source change when the observer moves outwards from the shower centre to the north and east, respectively. While the pulse amplitude drops quickly when one goes to the north, it stays fairly constant as one goes to the east. This effect was already discussed under the term ``reduced $\theta$'' in \citet{HuegeFalcke2003a}: The particle trajectory bends towards an observer in the east or west, so that he or she still sees the particle with a very small angle to the line of sight --- just during a different part of the trajectory. For the specific geometry chosen in this example, i.e. a particle pair starting off vertically downwards in the shower centre, the pulses even get broader as one goes outwards to the east, since one only sees an (asymmetric) half pulse in the centre and can see the full (symmetric) pulse only at considerable distance. All in all the behaviour is exactly as expected.

\begin{figure}
   \psfrag{t0ns}[c][t]{$t$~[ns]}
   \psfrag{E0muVpm}[c][t]{$\left|\vec{E}(t)\right|$~[$\mu$V m$^{-1}$]}
   \includegraphics[height=8.6cm,angle=270]{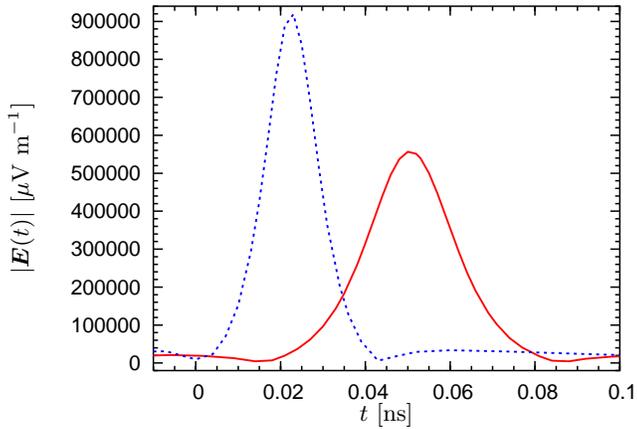}
   \caption{
   \label{fig:mcpulsesBfield}
   Individual particle pulses from a point-source shower as measured by an observer situated 995~m east of the shower centre. Solid: $B=0.3$~Gauss, dashed: $B=0.5$~Gauss. The time-integral over the pulses is constant.
   }
   \end{figure}

Another important consistency check is the dependence of the individual pulses on the magnetic field strength. Figure \ref{fig:mcpulsesBfield} shows the changes arising when the magnetic field is enhanced from 0.3~Gauss to 0.5~Gauss (both with 0$^{\circ}$ inclination, i.e.\ horizontal). The pulse gets stronger, yet at the same time shorter so that the integral over $E\,\mathrm{d}t$ stays constant (while the power an observer receives from an individual particle, i.e.\ the integral over $E^{2}\,\mathrm{d}t$, thus scales linearly with $B$), exactly as inferred from analytic calculations. This directly leads to an important result: The overall radio emission from an air shower cannot depend strongly on the magnetic field strength, as it is the sum of a great number of individual pulses, the integral of each of which is independent of the value of $B$. Likewise, asymmetries between north and south that are introduced by a realistically inclined magnetic field cannot be very strong.

The fact that the 0.5~Gauss pulse arrives earlier than the 0.3~Gauss pulse for an observer 995~m to the east of the centre is explained by the stronger curvature of the particle trajectory.

\subsection{Symmetry N-S and E-W} \label{sec:nssymmetry}

For a vertical air shower, the emitted radiation pattern must have a number of inherent symmetries. In particular, the east and west directions are completely equivalent (as long as particle track lengths for electrons and positrons are adopted as identical), so that the emission pattern must be symmetric in east and west. For a horizontal magnetic field, north and south must also be equal. The fact that these intrinsic symmetries are fulfilled as expected provides an important consistency check. Furthermore, one can use these symmetries to save computation time by calculating only a half- or even quarter-plane on the ground and then mirroring the results accordingly.

\subsection{Gridding algorithms} \label{sec:evalgriddingalgos}

As explained in section \ref{sec:griddingstrategy}, we have implemented a ``simple grid'' for low time-resolution calculations as well as an ``economic grid'' for cases in which the user is interested in the full time-resolution associated to the individual particle pulses.

\begin{figure}
   \psfrag{t0ns}[c][t]{$t$~[ns]}
   \psfrag{E0muVpm}[c][t]{$\left|\vec{E}(t)\right|$~[$\mu$V m$^{-1}$]}
   \includegraphics[height=8.6cm,angle=270]{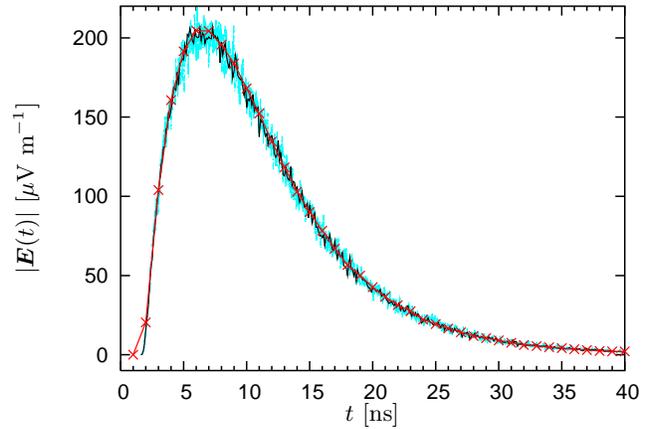}
   \caption{
   \label{fig:timerobustness}
   Raw pulse from a shower slice calculated with different gridding strategies and resolutions. Solid with points: simple grid $10^{-9}$~s, solid black: simple grid $10^{-10}$~s, light coloured: economic grid $10^{-12}$~s.
   }
   \end{figure}

The independence of the two algorithms allows a cross-check of the routines. Fig.\ \ref{fig:timerobustness} shows the raw pulse as calculated with the different gridding strategies at different time resolutions. The result is consistent between all three cases, which demonstrates that both algorithms work well. The ``simple grid'' proves to be very efficient at low time-resolutions (typically $\sim1$~ns). It is especially robust in the sense that it remains stable regardless of the specific resolution used. The ``economic grid'' on the other hand has to be set to a time-resolution high enough to resolve the individual particle pulses. Especially when particle energy distributions are switched on, introducing very high-energy particles with up to $\gamma=1000$, this strategy quickly becomes inefficient.

Additionally, low time-resolution ``simple grid'' data yields smoother pulses due to the time-averaging over the individual particle pulses, which allows precise calculations with fewer particles.

\subsection{Smart trajectory-sampling}

\begin{figure}
   \psfrag{EmaxmuVpm}[c][t]{Max$\left(\left|\vec{E}(t)\right|\right)$~[$\mu$V m$^{-1}$]}
   \includegraphics[height=8.6cm,angle=270]{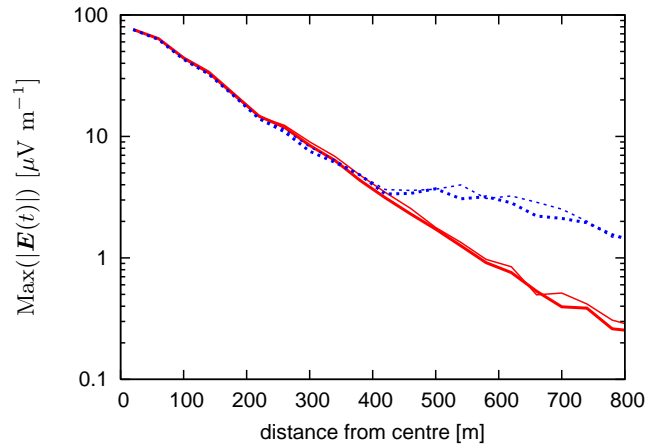}
   \caption{
   \label{fig:smartvsequis}
   Changes introduced by the smart sampling algorithm in the radial emission pattern of the rectangle-filtered maximum pulse amplitude for emission from the shower maximum. Thin lines: dense equidistant sampling, thick lines: smart sampling; solid: to the north, dashed: to the west.
   }
   \end{figure}

Figure \ref{fig:smartvsequis} demonstrates that the errors introduced by the smart sampling algorithm are only slight. At the same time, the algorithm allows a huge cut-down on computation time. The algorithm can, however, optionally be switched off for a more precise calculation.

\subsection{Cutting off $\gamma^{-1}$-cones} \label{sec:evalgammacones}

\begin{figure}
   \psfrag{EmaxmuVpm}[c][t]{Max$\left(\left|\vec{E}(t)\right|\right)$~[$\mu$V m$^{-1}$]}
   \includegraphics[height=8.6cm,angle=270]{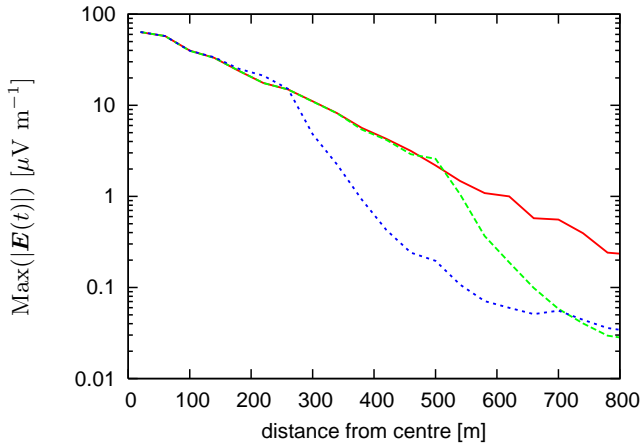}
   \caption{
   \label{fig:cuttingcones}
   Changes introduced by the cutting off of regions outside a few $\gamma^{-1}$-cones in the radial emission pattern of the frequency-filtered maximum pulse amplitude for emission from the shower maximum. Distance from the shower centre is to the north. Solid: no cutting, long dashed: cutting after $8\ \gamma^{-1}$, short dashed: cutting after $4\ \gamma^{-1}$
   }
   \end{figure}

As mentioned earlier, the discrete cutting off of radiation contributions outside the ground-trace of a few $\gamma^{-1}$-cones width can decrease the computation time enormously. Figure \ref{fig:cuttingcones}, however, demonstrates that this strategy is not suitable if one needs precision at the percent-level to be able to describe the emission pattern out to distances of several hundred metres from the shower centre.

The discrete cutting introduces ``breaks'' in the radial emission pattern exactly at the positions corresponding to the cutoff, i.e. at $\approx 530$~m in case of $8\ \gamma^{-1}$-cones and $\approx 270$~m in case of $4\ \gamma^{-1}$-cones for $\gamma=60$ particles at 4~km height. The effect is less strong in the east-west direction, but overall this strategy is disqualified for high-precision calculations.

The problems are resolved by the more sophisticated on-the-fly inactivation of ground-bins.

\subsection{Automatic ground-bin inactivation}

\begin{figure}
   \psfrag{EmaxmuVpm}[c][t]{Max$\left(\left|\vec{E}(t)\right|\right)$~[$\mu$V m$^{-1}$]}
   \includegraphics[height=8.6cm,angle=270]{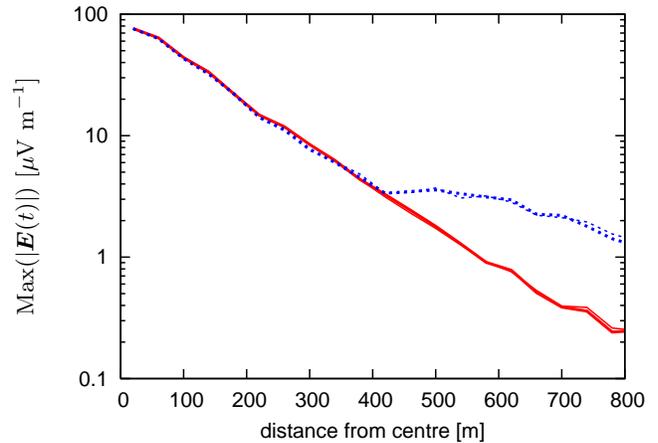}
   \caption{
   \label{fig:abivsnormal}
   Changes introduced to the radial emission pattern of the frequency-filtered maximum pulse amplitude for emission from the shower maximum by the automatic bin inactivation algorithm. Thin lines: no automatic bin inactivation, thick lines: automatic bin inactivation; solid: to the north, dashed: to the west.
   }
   \end{figure}

\begin{figure}
   \begin{center}
   \includegraphics[width=7.5cm]{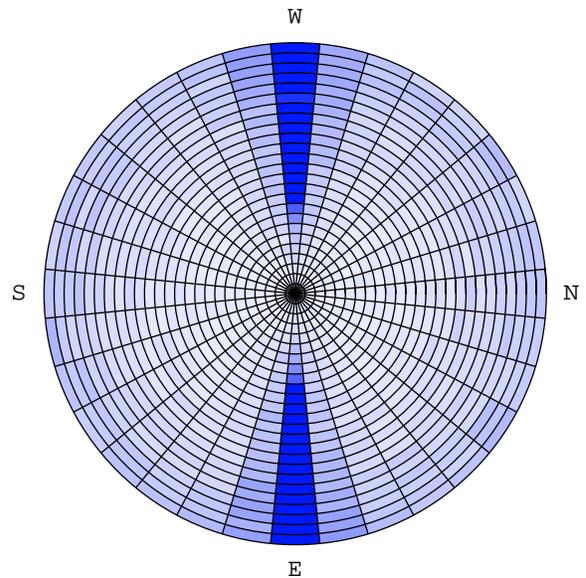}
   \end{center}
   \caption{
   \label{fig:passivationpar03}
   Automatic ground-bin inactivation sequence. Darker bins are set inactive later than lighter bins. The sequence propagates from the inside to the outside. The pattern is east-west and north-south symmetric as expected for a vertical shower and a horizontal magnetic field.
   }
   \end{figure}

Fig.\ \ref{fig:abivsnormal} demonstrates the stability of the automatic ground-bin inactivation algorithm. The calculations are as precise as those with a fixed high number of particles for all ground-bins, yet at the same time allow a much more efficient use of computing time and the definition of a user-specified precision goal. The deactivation sequence propagates from the inside to the outside as expected (see Fig.\ \ref{fig:passivationpar03}). It is also no surprise that the bins in the far east and west require the most computing time, as these are the regions that are influenced most by edge effects from trajectory cutoffs, different trajectory curvatures and the like. The automatic ground-bin inactivation strategy therefore turns out to be a very powerful and self-consistent technique to conduct extensive simulations with high precision.

\section{Emission from a shower slice} \label{sec:showerslice}

Similar to the analytical calculations described in \citet{HuegeFalcke2003a}, we first take a look at the emission from a single ``slice'' of the air shower. Throughout this section, we consider only the maximum of a vertical air shower induced by a $10^{17}$~eV primary particle, consisting of 10$^{8}$ charged particles at a height of 4~km. In a step by step analysis, we increase the complexity of the particle distributions and evaluate the changes introduced in the simulation results.

\subsection{Trajectory length effects} \label{sec:edgeeffects}

In a first step, we adopt the geomagnetic field parallel to the ground with a strength of 0.3~Gauss, as it is present at the equator. This is the same configuration that we used in the theoretical calculations. For the moment, we consider the simplified case of monoenergetic $\gamma \equiv 60$ particles.

The theoretical calculations carried out in \citet{HuegeFalcke2003a} were based on an analytical derivation of the spectra of individual particles on circular trajectories. This derivation makes the implicit assumption that the trajectory is always symmetric with respect to the point in which the minimum angle $\theta$ to the observer's line of sight is reached. In other words, edge effects arising from the cutting off of the finite trajectories are not taken into account.

\begin{figure}
   \psfrag{EmaxmuVpm}[c][t]{Max$\left(\left|\vec{E}(t)\right|\right)$~[$\mu$V m$^{-1}$]}
   \includegraphics[height=8.6cm,angle=270]{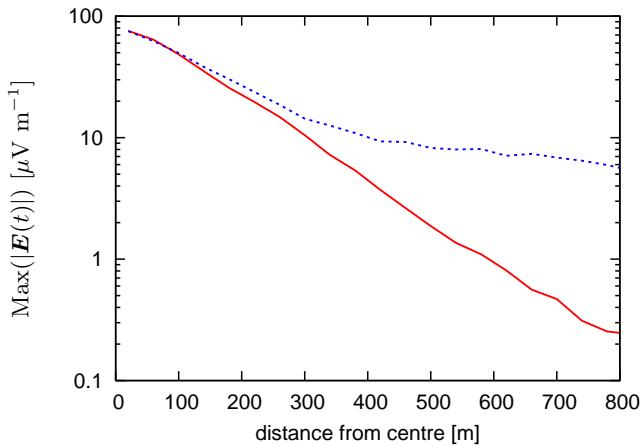}
   \caption{
   \label{fig:longtrajecasymmetry}
   Radial dependence of the maximum rectangle-filtered pulse amplitude for emission from the shower maximum in the north (solid) and west (dashed) direction in case of constant and long particle trajectories ($\lambda \equiv 100$~g~cm$^{-2}$, no edge effects at high distances from the shower centre).
   }
   \end{figure}

These edge effects, however, turn out to significantly shape the radial emission pattern of the radiation. Let us first consider the case of trajectories which are long enough to not produce significant edge effects for observers far away from the shower centre by adopting a trajectory length of 100~g~cm$^{-2}$. The result is depicted in Fig.\ \ref{fig:longtrajecasymmetry}. As expected, it is very similar to the theoretical prediction for the ``reduced $\theta$'' case presented in Fig.\ 14 of \citet{HuegeFalcke2003a}. In this scenario, one would expect a significant asymmetry between the north-south and east-west direction, which would obviously be a very useful observable.

\begin{figure}
   \psfrag{EmaxmuVpm}[c][t]{Max$\left(\left|\vec{E}(t)\right|\right)$~[$\mu$V m$^{-1}$]}
   \includegraphics[height=8.6cm,angle=270]{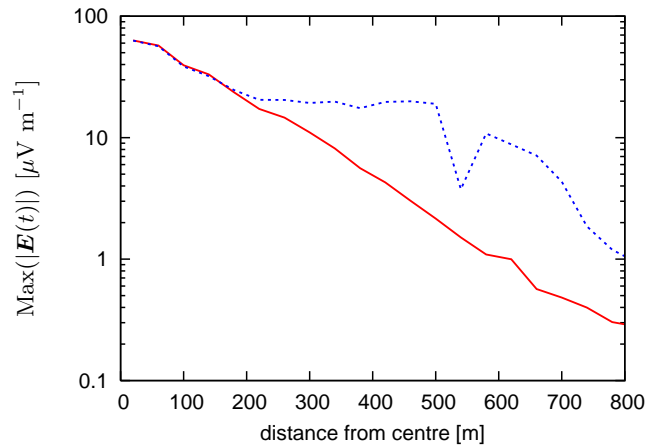}
   \caption{
   \label{fig:maxampasymmetryconsttraj}
   Same as Fig.\ \ref{fig:longtrajecasymmetry} for $\lambda \equiv 40$~g~cm$^{-2}$. See text for explanation of the ``kink'' at $\sim$540~m.
   }
   \end{figure}

\begin{figure}
   \psfrag{Eew0muVpm}[c][t]{$E_{\mathrm{EW}}(t)$~[$\mu$V m$^{-1}$]}
   \includegraphics[height=8.6cm,angle=270]{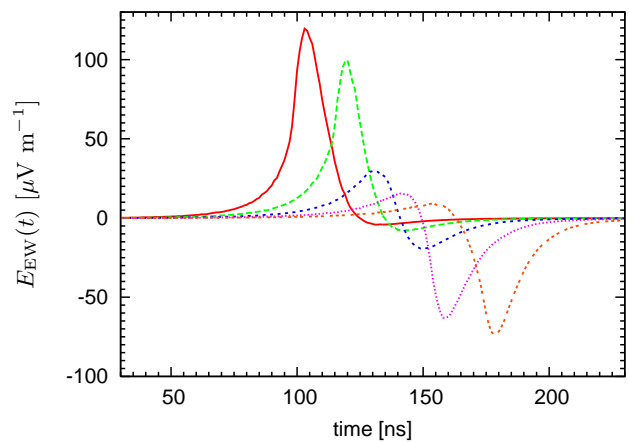}
   \caption{
   \label{fig:polaritychange}
   Time-dependence of the raw pulses originating from the shower maximum as observed by an observer at (from left to right) 460~m, 500~m, 540~m, 580~m and 620~m to the west from the shower centre. See text for explanation of the ``polarity change''.
   }
   \end{figure}

What happens, however, if one adopts more realistic particle track lengths? Let us first consider a constant trajectory length of 40~g~cm$^{-2}$, which is approximately the free path length (equal to one radiation length) of electrons and positrons in air \citep{Allan1971}. In this scenario, edge effects strongly shape the emission at high distances in the east-west direction as shown in Fig.\ \ref{fig:maxampasymmetryconsttraj}. An unusual ``kink'' appears at $\sim$540~m. This is not a numerical glitch, but an interference effect arising when the observer stops to see the main (positive) peak of the individual particle pulses due to the edge effects. He then only receives the initial (negative) contribution of the electric field pulse. This effectively causes a polarity change in the raw pulses as shown in Fig.\ \ref{fig:polaritychange}, accompanied by a temporary drop in the filtered pulse amplitude.

\begin{figure}
   \psfrag{EmaxmuVpm}[c][t]{Max$\left(\left|\vec{E}(t)\right|\right)$~[$\mu$V m$^{-1}$]}
   \includegraphics[height=8.6cm,angle=270]{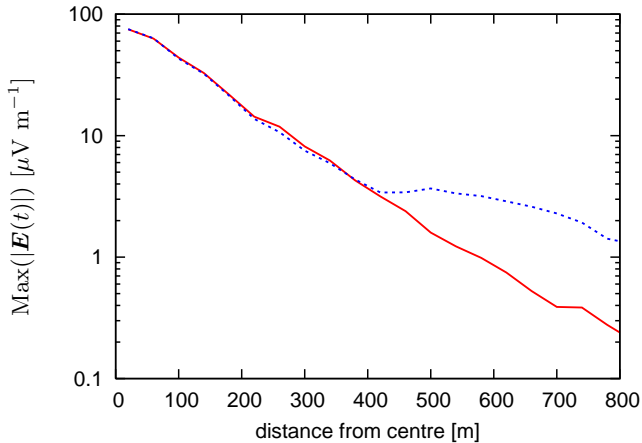}
   \caption{
   \label{fig:maxampasymmetrystattraj}
   North (solid) vs. west (dashed) asymmetry in the maximum filtered pulse amplitude for emission from the shower maximum in case of statistically distributed track-lengths. The asymmetry is washed out up to high distances.
   }
   \end{figure}

Finally, we change to a realistic exponential distribution of track lengths with a mean of $\lambda = 40$~g~cm$^{-2}$,
\begin{equation} \label{eqn:parlengthdistribution}
p(X) = p_{0}\ \exp\left(-\frac{X}{\lambda}\right).
\end{equation}
As can be seen in Fig.\ \ref{fig:maxampasymmetrystattraj}, the asymmetry between north-south and east-west direction is now washed out up to high distances. (In fact, it will be washed out almost completely once the integration over the shower evolution is taken into account.)

Apart from the regions far from the shower centre, edge effects also occur in the centre region due to the instantaneous starting of the trajectories. These effect are discussed in section \ref{sec:slicecomparison}.

The significance of the trajectory length effects already illustrates the importance of adopting as realistic properties for the particle distributions as possible, a goal that could not be reached with analytic calculations alone. We retain the realistic statistical distribution of track lengths for the following discussions of the magnetic field and energy distribution effects.

\subsection{Magnetic field dependence}

\begin{figure}
   \psfrag{EmaxmuVpm}[c][t]{Max$\left(\left|\vec{E}(t)\right|\right)$~[$\mu$V m$^{-1}$]}
   \includegraphics[height=8.6cm,angle=270]{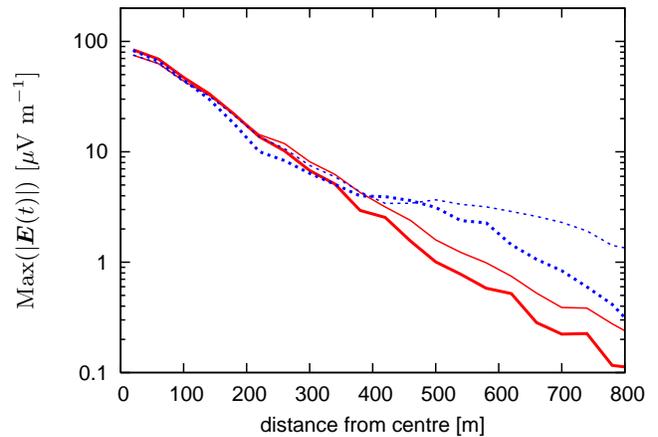}
   \caption{
   \label{fig:par03vsreal05asymmetry}
	Changes to the north (solid) and west (dashed) radial emission patterns for emission from the shower maximum when going from a 0.3~Gauss horizontal magnetic field (thin lines) to a 70$^{\circ}$ inclined 0.5~Gauss magnetic field (thick lines).
   }
   \end{figure}

We now make the change from an equatorial 0.3~Gauss horizontal magnetic field to the 0.5~Gauss 70$^{\circ}$ inclined magnetic field present in central Europe. As explained in Sec.\ \ref{sec:individualpulses}, the influence of the magnetic field on the integrated shower pulse should not be very significant. This is confirmed by Fig.\ \ref{fig:par03vsreal05asymmetry}. Although the projected magnetic field strength drops from 0.3~Gauss to 0.17~Gauss, the emission pattern only changes very slightly. In particular, the amplitude level stays almost constant. Interestingly, the north-south vs.\ east-west asymmetry is washed out even further.

   \begin{figure*}
   \includegraphics[width=4.1cm,angle=270]{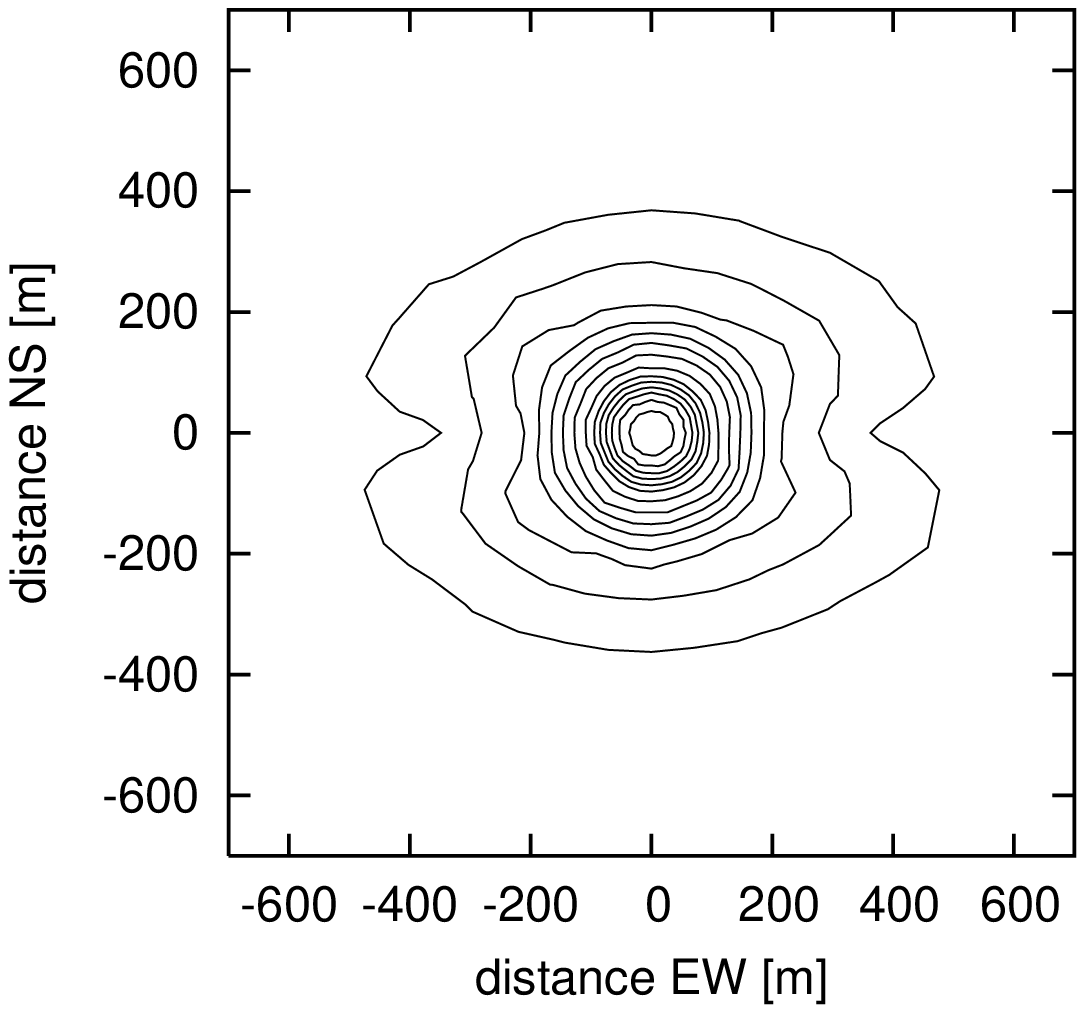}
   \includegraphics[width=4.1cm,angle=270]{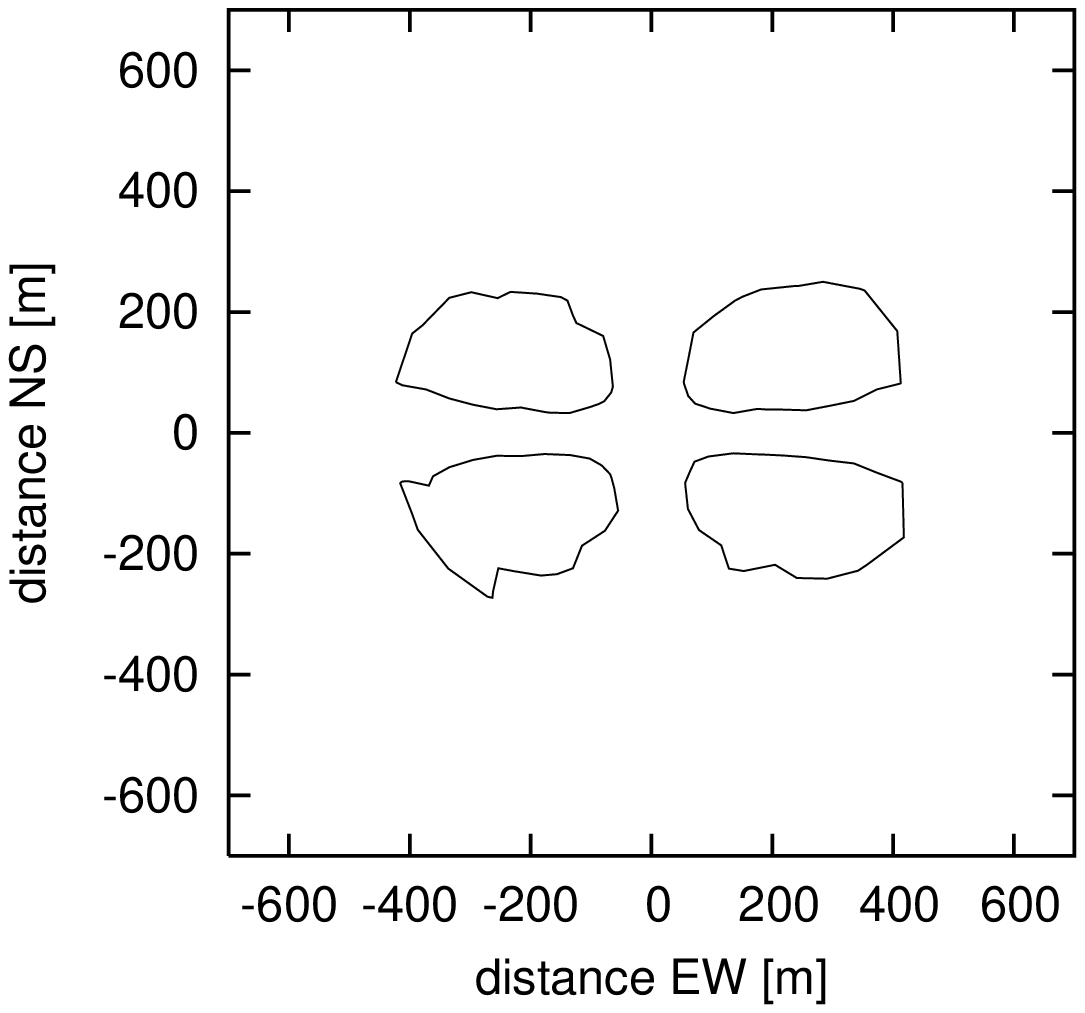}
   \includegraphics[width=4.1cm,angle=270]{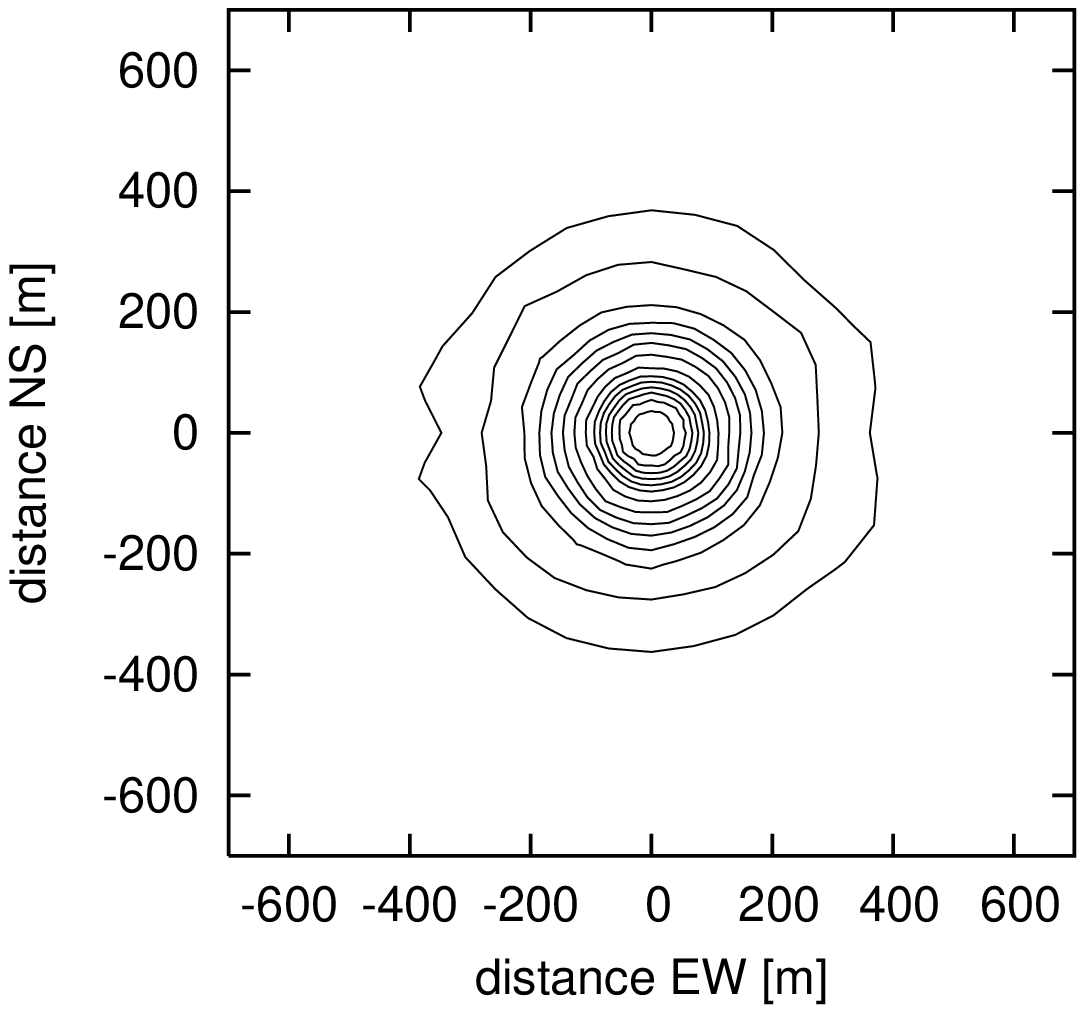}
   \includegraphics[width=4.1cm,angle=270]{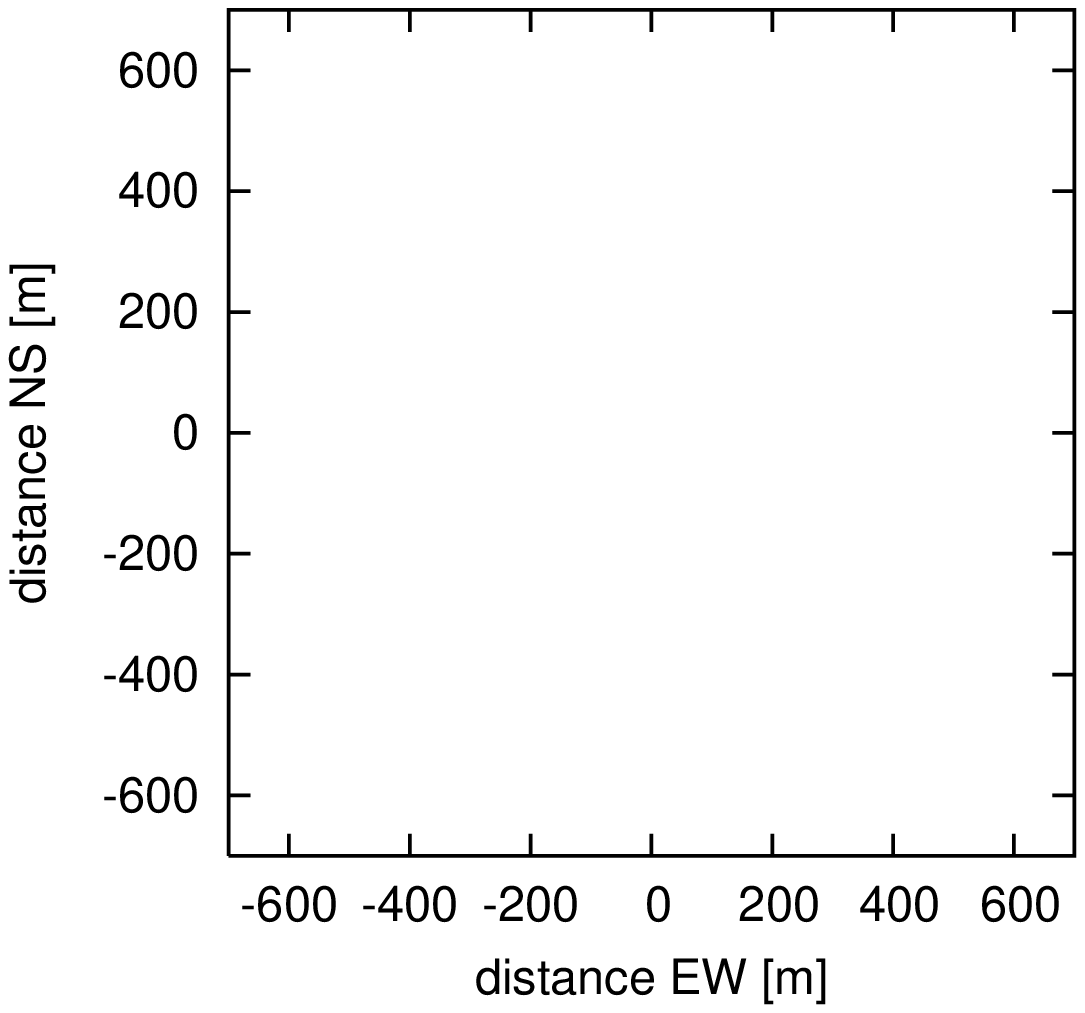}\\
   \includegraphics[width=4.1cm,angle=270]{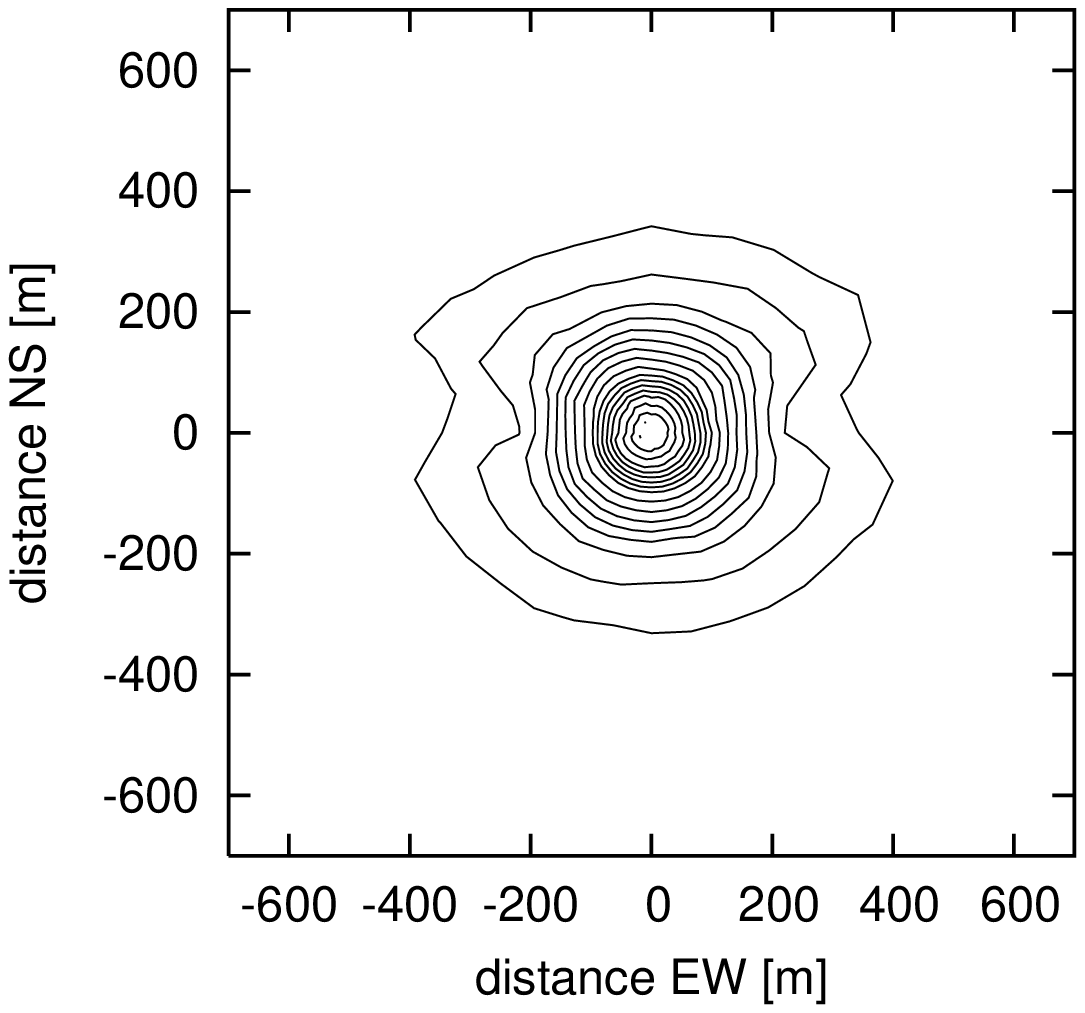}
   \includegraphics[width=4.1cm,angle=270]{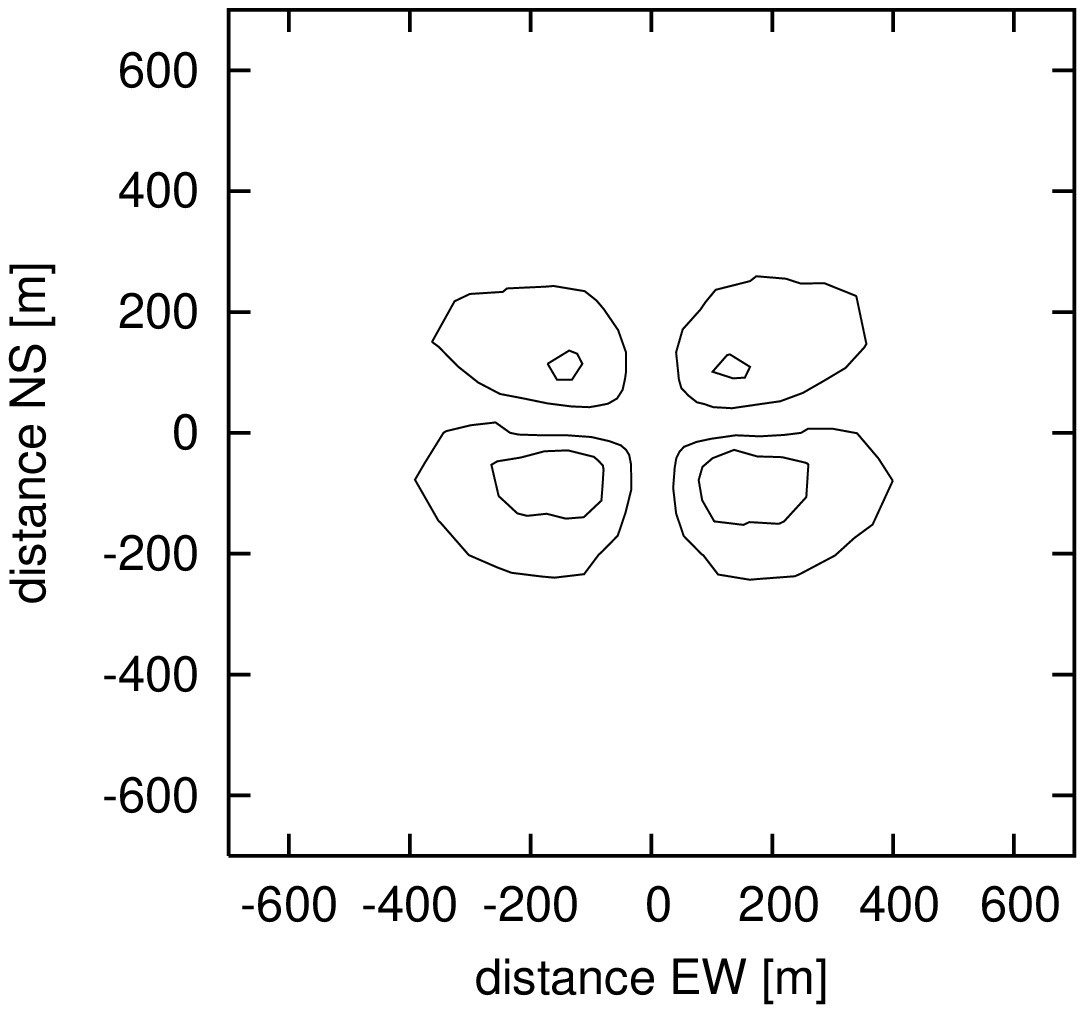}
   \includegraphics[width=4.1cm,angle=270]{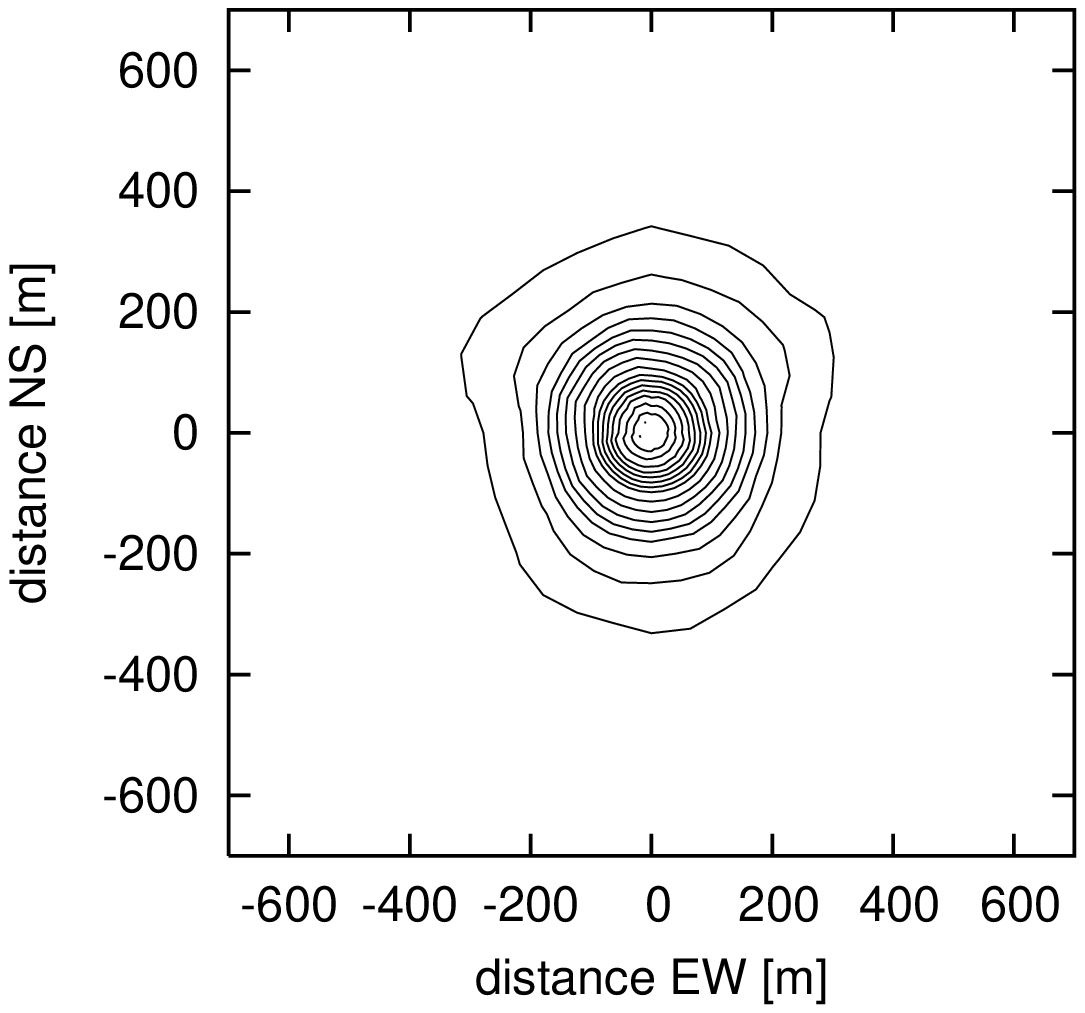}
   \includegraphics[width=4.1cm,angle=270]{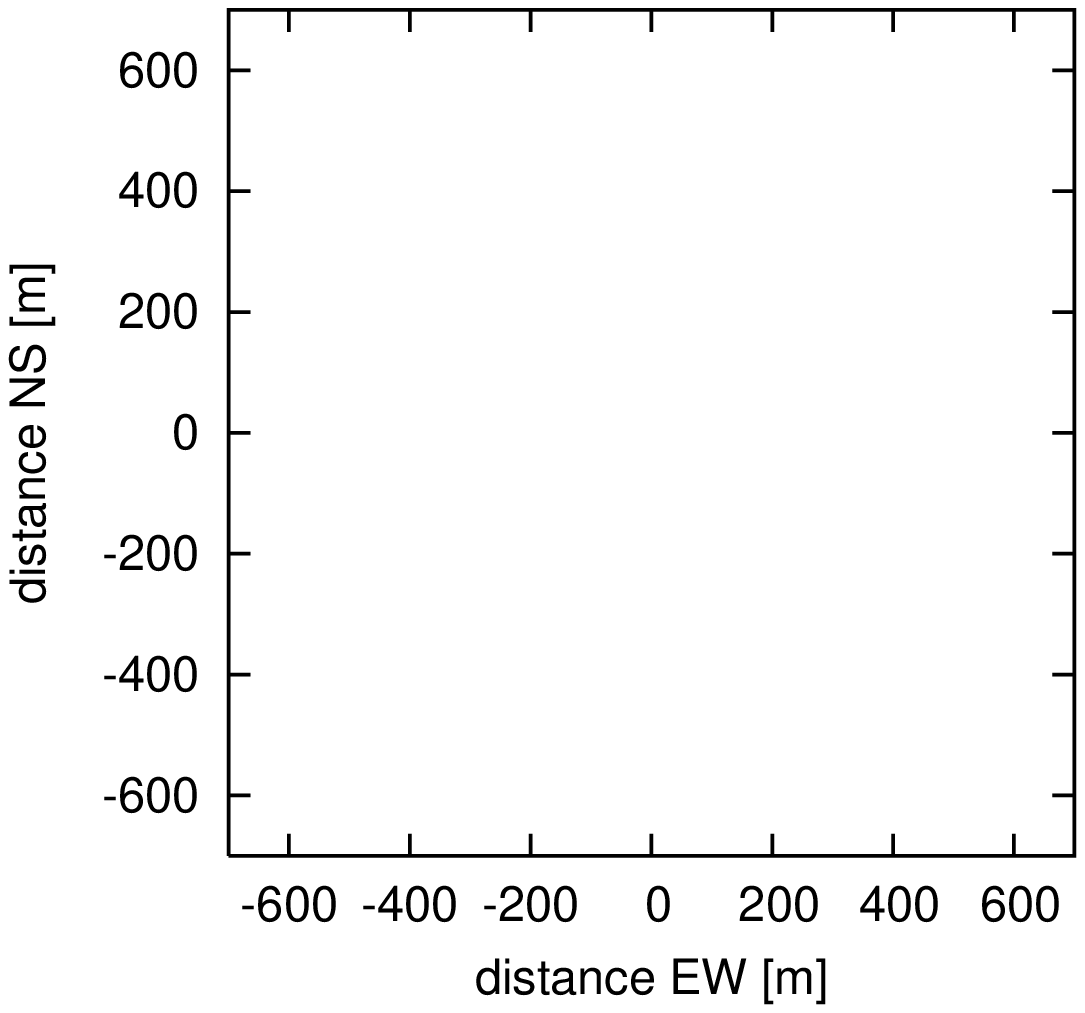}
   \caption{
   \label{fig:par03real05contours}
   Contour plots of the 40--160~MHz rectangle-filtered maximum pulse amplitude for emission from the shower maximum in case of a horizontal 0.3~Gauss magnetic field (upper panel) and a 70$^{\circ}$ inclined 0.5~Gauss magnetic field (lower panel) and $\gamma\equiv 60$ particles.  Contour levels are 5~$\mu$V~m$^{-1}$ apart. From left to right: total electric field strength, north-south polarisation component, east-west polarisation component, vertical polarisation component.
   }
   \end{figure*}

\begin{figure}
   \begin{center}
   \includegraphics[width=7.5cm]{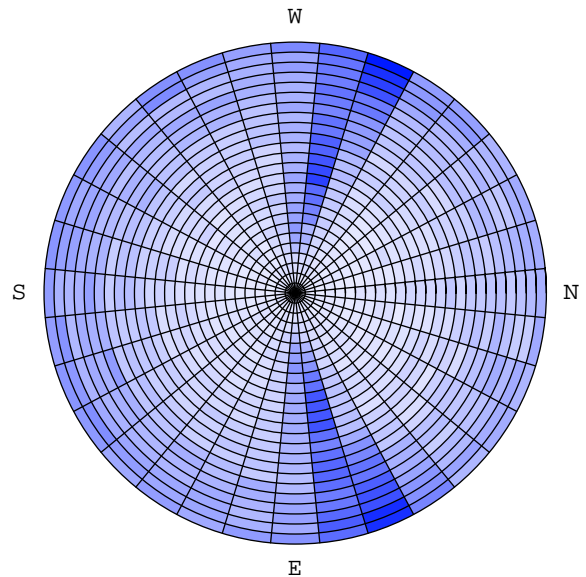}
   \end{center}
   \caption{
   \label{fig:passivationreal05}
   Automatic ground-bin inactivation sequence in case of 70$^{\circ}$ inclined 0.5~Gauss magnetic field. The north-south symmetry is broken as expected, cf.\ Fig.\ \ref{fig:passivationpar03}.
   }
   \end{figure}

On the other hand, the inclination of the magnetic field breaks the intrinsic north-south symmetry of the emission pattern. This is visible in the north-south and east-west polarisation components of the contour plots shown in Fig.\ \ref{fig:par03real05contours}. Similarly, the symmetry of the ground-bin inactivation sequence is broken as shown in Fig.\ \ref{fig:passivationreal05}. The effect is, however, only weak and there is still no asymmetry if one considers the total electric field strength rather than a specific polarisation direction.

Note that there is no significant emission in the centre region of the north-south polarisation component (second column of Fig.\ \ref{fig:par03real05contours}). This demonstrates that the numerical cancellation of north-south polarised radiation components from the electrons and positrons (labelled $A_{\perp}$ in \citet{HuegeFalcke2003a}) indeed works correctly in the MC code and that it was justified to neglect this component in the earlier theoretical calculations at least for the primarily important centre region.

In summary, the magnetic field effects are weak, but non-trivial.

\subsection{Energy distribution effects}

\begin{figure}
   \psfrag{EmaxmuVpm}[c][t]{Max$\left(\left|\vec{E}(t)\right|\right)$~[$\mu$V m$^{-1}$]}
   \includegraphics[height=8.6cm,angle=270]{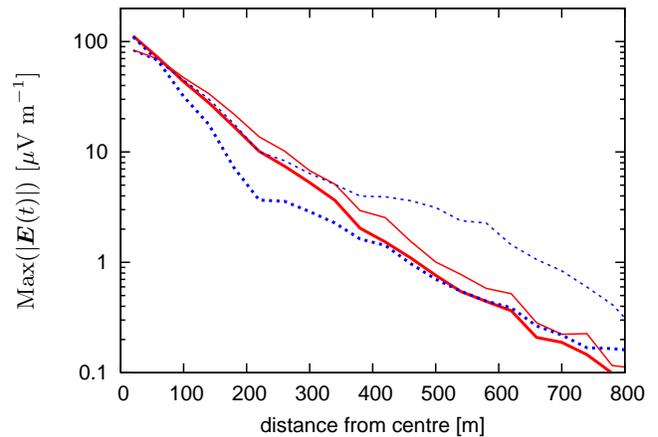}
   \caption{
   \label{fig:enerdistrivsmonoener}
   Changes introduced when switching from monoenergetic $\gamma \equiv 60$ particles (thin lines) to a broken power-law peaking at $\gamma = 60$ (thick lines) for emission from the shower maximum. Solid: to the north, dashed: to the west. See text for explanation of the drop at $\sim$200~m.
   }
   \end{figure}

In the theoretical calculations, a change from monoenergetic $\gamma \equiv 60$ electrons to a broken power-law peaking at $\gamma=60$ introduced only minor changes, namely a slight redistribution of radiation from medium distances to the centre region (due to the high-energy particles) and high distances (due to the low-energy particles). The changes introduced by the energy distribution are bound to be more complex in the MC simulations, as they are heavily intertwined with the edge effects discussed before. Nonetheless, a similar redistribution of radiation from the medium scales to the innermost centre region is observable in the MC simulations as shown in Fig.\ \ref{fig:enerdistrivsmonoener}.

The drop visible in the west direction at $\sim$200~m arises due to a transition from a dominating east-west polarisation component to a dominating north-south polarisation component in combination with some resolving out of pulses due to the filter bandwidth used. (Fig.\ \ref{fig:par03real05contours} demonstrates that the north-south and east-west polarisation components become comparable at these distances.)

It should be pointed out, however, that the strength of the effects introduced by the choice of a specific energy distribution seem misleadingly strong when one considers the emission from a single shower slice alone. Once the integration over the shower evolution as a whole is performed, most effects (including the drop at $\sim$200~m) are again washed out almost completely and the influence of the specific choice of energy distribution becomes very weak. For the moment, we therefore continue to use the broken power-law distribution that we adopted in the theoretical calculations rather than implementing a more realistic distribution such as the one given by \citet{NerlingEngelGuerard2003}.

   \begin{figure}
   \includegraphics[height=8.6cm,angle=270]{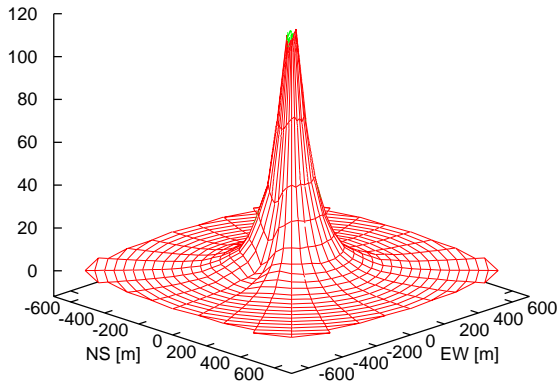}
   \caption{
   \label{fig:surfacereal05}
   Pattern of the maximum filtered electric field amplitude (in $\mu$V~m$^{-1}$) emitted by the maximum of a $10^{17}$~eV air shower consisting of $10^8$ particles at a height of 4~km  for a 0.5~Gauss 70$^{\circ}$ inclined magnetic field, a broken power-law particle energy distribution and a statistical distribution of track lengths.
   }
   \end{figure}

The result that we have reached so far is the emission from the maximum of a $10^{17}$~eV air shower consisting of $10^{8}$ electrons and positrons at a height of 4~km, taking into account adequate spatial, energy and trajectory length distributions and a magnetic field as present in central Europe. It is illustrated once more as a surface plot in Fig.\ \ref{fig:surfacereal05}.

\subsection{Comparison with theoretical calculations} \label{sec:slicecomparison}

As shown in Fig.\ \ref{fig:mcanalyticpulse} the individual particle pulses calculated with our MC code are consistent with the results of the analytic calculations. The analytic pulses are symmetric because the particle trajectories are implicitly adopted symmetric to the point of smallest angle to the line of sight --- the analytic calculations do not consider edge effects associated with the finite lengths of the trajectories. When one takes into account the cutting off of the trajectories correctly in the MC calculations, an observer in the centre region only sees half of the symmetric pulse for each individual particle (cf.\ Fig.\ \ref{fig:mcpulsesE}).

In section \ref{sec:individualpulses} we discussed that the time integral over the individual particle pulses is the quantity relevant for the overall pulse as integrated over the shower as a whole. For simple geometries such as a point source or a line charge, the change from the symmetric to the half pulses should therefore produce a drop in the overall pulse amplitude (and the overall pulse spectrum) by a factor $\sim2$. (For more complex shower geometries or at higher distances from the shower centre the changes introduced by the edge effects are non-trivial but still important as already discussed in section \ref{sec:edgeeffects}.)

\begin{figure}
   \psfrag{Eomegaew0muVpmpMHz}[c][t]{$E_{\mathrm{EW}}(\vec{R},2\pi\nu)$~[$\mu$V~m$^{-1}$~MHz$^{-1}$]}   
   \psfrag{nu0MHz}[c][t]{$\nu$~[MHz]}
   \includegraphics[height=8.6cm,angle=270]{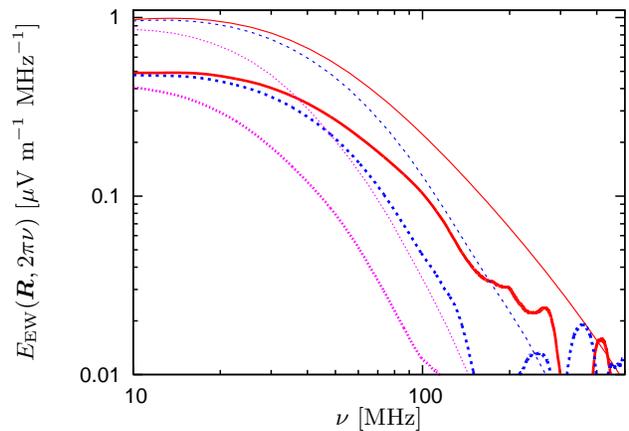}
   \caption{
   \label{fig:speccomp_monoener_slice_factor1.0}
   Spectra emitted by the maximum of a $10^{17}$~eV vertical air shower consisting of $10^{8}$ particles with $\gamma \equiv 60$ at a height of 4~km, long (constant 100~g~cm$^{-2}$), but not symmetric, particle trajectories and horizontal 0.3~Gauss magnetic field. Thin lines: analytic calculations of \protect\citet{HuegeFalcke2003a}, thick lines: these MC simulations. Solid: 20~m, dashed: 100~m, dotted: 260~m to north from shower centre.
   }
   \end{figure}

\begin{figure}
   \psfrag{Eomegaew0muVpmpMHz}[c][t]{$E_{\mathrm{EW}}(\vec{R},2\pi\nu)$~[$\mu$V~m$^{-1}$~MHz$^{-1}$]}   
   \psfrag{nu0MHz}[c][t]{$\nu$~[MHz]}
   \includegraphics[height=8.6cm,angle=270]{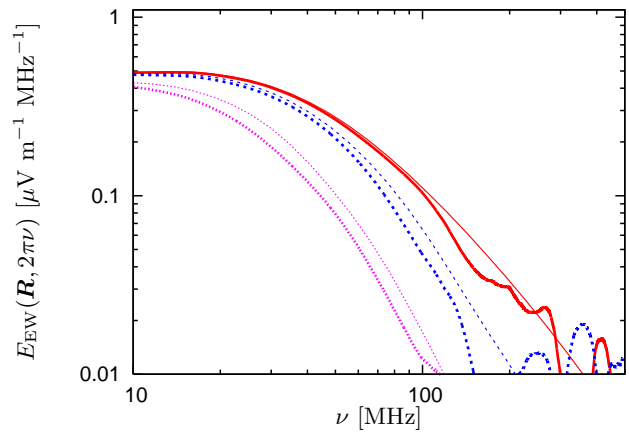}
   \caption{
   \label{fig:speccomp_monoener_slice_factor2.0}
   Same as Fig.\ \ref{fig:speccomp_monoener_slice_factor1.0} but scaling down the analytic results by a factor of two.
   }
   \end{figure}

\begin{figure}
   \psfrag{Eomegaew0muVpmpMHz}[c][t]{$E_{\mathrm{EW}}(\vec{R},2\pi\nu)$~[$\mu$V~m$^{-1}$~MHz$^{-1}$]}   
   \includegraphics[height=8.6cm,angle=270]{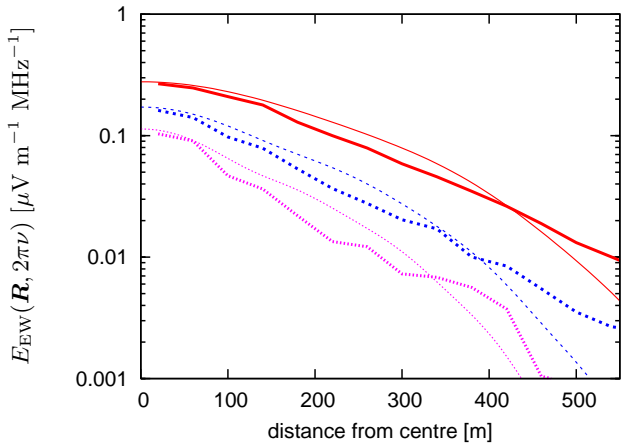}
   \caption{
   \label{fig:radialcomp_monoener_slice_factor2.0}
   Radial dependence of the emission for the same scenario as in Fig.\ \ref{fig:speccomp_monoener_slice_factor1.0}. Thin lines: analytic calculations of \protect\citet{HuegeFalcke2003a} scaled down by a factor of two, thick lines: these MC simulations. Solid: $\nu=50$~MHz, dashed: $\nu=75$~MHz, dotted: $\nu=100$~MHz. Distance is to north from shower centre.
   }
   \end{figure}

Fig.\ \ref{fig:speccomp_monoener_slice_factor1.0} shows the direct comparison between the analytic and MC simulated spectra emitted by a slice of monoenergetic $\gamma \equiv 60$ particles for the case of a horizontal $0.3$~Gauss magnetic field and long (constant 100~g~cm$^{-2}$), but not symmetric, particle trajectories. (Because the north-south polarisation component was neglected in the analytic calculations, we hereafter directly compare the east-west polarisation component.) This scenario allows a very direct comparison between the analytic and MC calculations, the only major difference being the edge effects introduced due to the non-symmetric trajectories. Consequently, the analytic spectra lie a factor $\sim2$ above the MC results. Scaling down the analytic results by a factor of two shows indeed a very good agreement between the analytic and MC results as seen in Fig.\ \ref{fig:speccomp_monoener_slice_factor2.0}. The radial dependence of the emission pattern, shown in Fig.\ \ref{fig:radialcomp_monoener_slice_factor2.0} also shows good agreement between the MC and analytic calculations when one accounts for the systematic factor of two.

In a next step, we switch on the statistical distribution of trajectory lengths with a realistic mean free path length of 40~g~cm$^{-2}$, adopt again a broken power-law distribution of particle energies and change to the realistic 70$^{\circ}$ inclined 0.5~Gauss magnetic field present in Central Europe. In the analytic calculations, we switch on the broken power-law distribution, but cannot take into account the track length effects or the inclined magnetic field. Obviously, the results produced by the MC simulations in this scenario therefore cannot be expected to reproduce the analytic results equally well.

   \begin{figure}
   \psfrag{Eomegaew0muVpmpMHz}[c][t]{$E_{\mathrm{EW}}(\vec{R},2\pi\nu)$~[$\mu$V~m$^{-1}$~MHz$^{-1}$]}   
   \psfrag{nu0MHz}[c][t]{$\nu$~[MHz]}
   \includegraphics[height=8.6cm,angle=270]{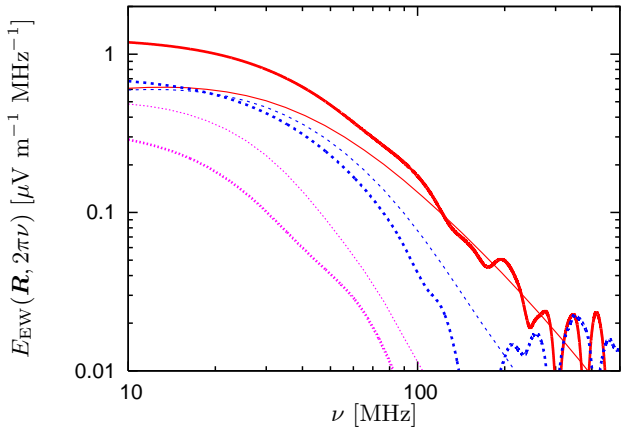}
   \caption{
   \label{fig:speccomp_enerdistri_stattraj_real05_slice_factor2.0}
   Spectra emitted by the maximum of a $10^{17}$~eV vertical air shower consisting of $10^{8}$ particles with broken power-law energy distribution at a height of 4~km, statistically distributed particle trajectories with 40~g~cm$^{-2}$ mean path length and 70$^{\circ}$ inclined 0.5~Gauss magnetic field. Thin lines: analytic calculations of \protect\citet{HuegeFalcke2003a} scaled down by a factor of two, thick lines: these MC simulations. Solid: 20~m, dashed: 100~m, dotted: 260~m to north from shower centre.
   }
   \end{figure}

   \begin{figure}
   \psfrag{Eomegaew0muVpmpMHz}[c][t]{$E_{\mathrm{EW}}(\vec{R},2\pi\nu)$~[$\mu$V~m$^{-1}$~MHz$^{-1}$]}   
   \includegraphics[height=8.6cm,angle=270]{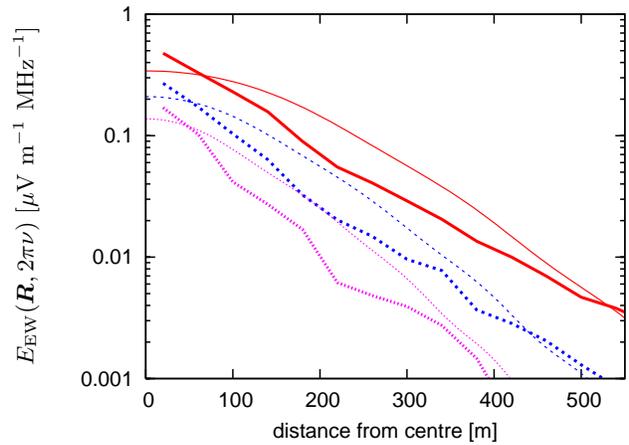}
   \caption{
   \label{fig:radialcomp_enerdistri_stattraj_real05_slicefactor2.0}
   Radial dependence of the emission for the same scenario as in Fig.\ \ref{fig:speccomp_enerdistri_stattraj_real05_slice_factor2.0}. Thin lines: analytic calculations of \protect\citet{HuegeFalcke2003a} scaled down by a factor of two, thick lines: these MC simulations. Solid: $\nu=50$~MHz, dashed: $\nu=75$~MHz, dotted: $\nu=100$~MHz. Distance is to north from shower centre.
   }
   \end{figure}

A direct comparison between the spectral dependences predicted by the analytic and MC calculations is shown in Fig.\ \ref{fig:speccomp_enerdistri_stattraj_real05_slice_factor2.0}, keeping the down-scaling of the analytic results by a factor of two to compensate for the symmetric trajectories. As discussed earlier, switching on the energy distribution redistributes flux from medium scales to the centre region. Correspondingly, the spectrum at 20~m distance shifts to higher amplitudes. The effect is stronger in the MC calculations than in the analytics. The MC 100~m spectrum fits well with the analytic results, whereas the MC 260~m spectrum lies at lower amplitudes than in the analytics: The radial dependence now falls off much steeper in the MC as compared to the analytic results, as is also visible in Fig.\ \ref{fig:radialcomp_enerdistri_stattraj_real05_slicefactor2.0}. Overall, the radial dependence follows a very much exponential decay. (We will carry out a detailed analysis of the functional form of different properties of the radio emission in a later paper.) It does not exhibit the prominent plateaus visible in the centre regions of the analytic calculations. Considering the much higher precision of the MC simulations and the neglect of edge effects and statistical trajectory lengths in the analytic calculations, however, we consider the agreement quite acceptable.


\section{Emission from an integrated shower}

After having analysed the emission from a shower slice, the next step in our analysis now is to perform the integration over the air shower as a whole.

\subsection{Integration over shower evolution}

In the theoretical calculations we performed in \citet{HuegeFalcke2003a}, the integration over the shower evolution was carried out in a somewhat simplified way: The shower evolution was discretised into slices of independent generations of particles, spaced apart by one radiation length each. The overall emission was then superposed as the sum of the radiation from all these slices. Although this strategy should allow a good estimate of the emission from the complete shower, it has at least two problems: First, the total number of particles in the shower --- and thus the total field amplitude --- is directly influenced by the scale introduced through the spacing of the slices. A denser or wider spacing directly leads to higher or lower emission levels, respectively. Although the radiation length is the logical choice for this scale, a ``scale-free'' approach would be a better choice. Second, our theoretical calculations, strictly speaking, are only valid in the far field. Consequently, the emission from slices close to the ground, especially for high-energy showers, cannot be taken into account with the desired precision.

Both these pitfalls no longer pose a problem in the MC simulations: No far-field approximations at all were applied in the MC calculations, and the continuous evolution of the shower is correctly taken into account in the creation of particles according to the corresponding probability distribution function.

For the shower profile we use the Greisen parametrisation \citep{Greisen1960} that we already adopted in \citet{HuegeFalcke2003a}:
\begin{equation}\label{eqn:Nelongitudinal}
N(s)=\frac{0.31 \exp\left[\frac{X_{\mathrm{m}}}{X_{0}}\frac{2-3 \ln s}{3/s-1}\right]}{\sqrt{X_{\mathrm{m}}/X_{0}}},
\end{equation}
where the (theoretical) position of the shower maximum $X_{\mathrm{m}}$ is given by
\begin{equation} \label{eqn:xmax}
X_{\mathrm{m}} = X_{0}  \ln\left(E_{\mathrm{p}}/E_{\mathrm{crit}}\right),
\end{equation}
$X_{0}=36.7$~g~cm$^{-2}$ denotes the electron radiation length in air, $E_{\mathrm{crit}}=86$~MeV corresponds to the threshold energy where ionisation losses equal radiation losses for electrons moving in air, and $E_{\mathrm{p}}$ specifies the primary particle energy. Equation (\ref{eqn:xmax}) predicts the position of the shower maximum for a purely electromagnetic cascade, in which the shower age as a function of atmospheric depth then corresponds to
\begin{equation}
s(X)=\frac{3X}{X+2X_{\mathrm{m}}}.
\end{equation}
Obviously, this parametrisation of the shower age has to be modified to adequately describe the evolution of the hadronic air showers in our MC simulations. We choose to manually set the depth of the shower maximum to an empirical value $X_{\mathrm{m,e}}$ as a function of primary particle energy and shower inclination, e.g.\ $X_{\mathrm{m,e}}\sim 630$~g~cm$^{-2}$ for our typical $10^{17}$~eV vertical air shower. The shower age is then adopted as
\begin{equation}
s(X)=\frac{3X}{X+2X_{\mathrm{m,e}}},
\end{equation}
whereas eq.\ (\ref{eqn:Nelongitudinal}) is left unchanged, i.e.\ retaining the theoretically motivated value for $X_{\mathrm{m}}$.\footnote{A more realistic set of parameters for application of the Greisen function to hadronic showers in the energy range between $10^{17}$ and $10^{18}$~eV was established in \protect\citet{AbuZayyadBelovBird2001}. For the moment, however, we retain the parametrisation as stated above to allow a better comparison with our earlier theoretical calculations.}

\begin{figure}
   \psfrag{xdepth}[c][t]{$X$ [g~cm$^{-2}$]}   
   \psfrag{numinje6}[c][t]{$I(X)$ [10$^{6}$~g$^{-1}$~cm$^{2}$]}   
   \psfrag{numee7}[c][t]{$N(X)$ [10$^{7}$]}   
   \includegraphics[height=8.6cm,angle=270]{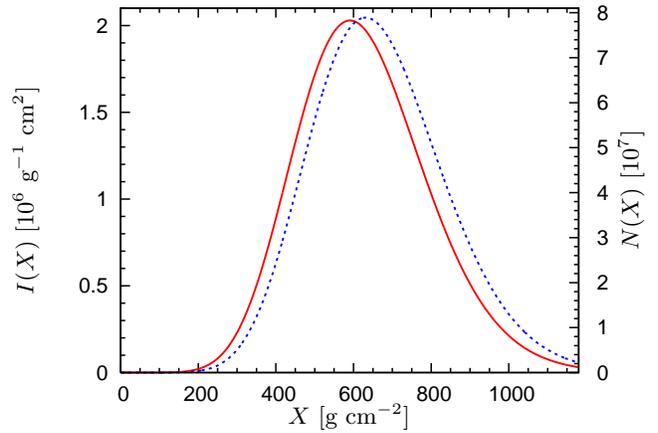}
   \caption{
   \label{fig:numvsinj}
   Comparison of $I(X)$ (solid) and $N(X)$ (dashed) as a function of atmospheric depth $X$ for a vertical 10$^{17}$~eV shower with $X_{\mathrm{m,e}}=631$~g~cm$^{-2}$ and $\lambda=40$~g~cm$^{-2}$.
   }
   \end{figure}

Eq.\ (\ref{eqn:Nelongitudinal}) denotes the integrated number of electronic particles that a detector positioned at atmospheric depth $X$ measures as the shower sweeps through it. This number is {\em{not}} equal to the number of particles $I(X)$ that are ``injected'' at that atmospheric depth, the quantity we need to describe the probability distribution function for the creation of particles. The two quantities are directly related via the path length distribution of the particles. For an exponential path length distribution with mean free path length $\lambda$ as given in eq.\ (\ref{eqn:parlengthdistribution}), the injection function is given by
\begin{equation} \label{eqn:inject}
I(X) = \frac{\mathrm{d}N}{\mathrm{d}X} + \frac{N}{\lambda}.
\end{equation}
As demonstrated in Fig.\ \ref{fig:numvsinj}, $I(X)$ closely follows the form of $N(X)$ with an offset of $\sim\lambda$ to lower $X$ values, i.e.
\begin{equation}
I(X) \approx N(X+\lambda)/\lambda.
\end{equation}

\begin{figure}
   \includegraphics[height=8.6cm,angle=270]{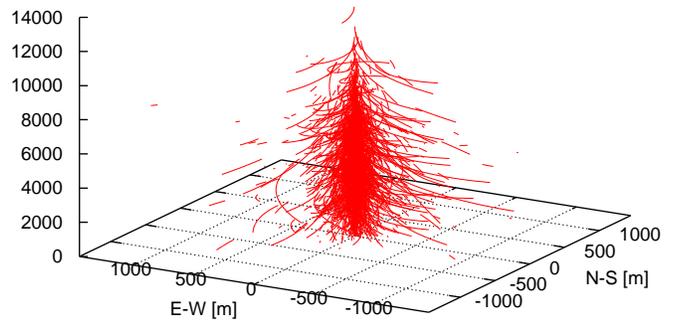}
   \caption{
   \label{fig:trajectrace}
   Trace of the trajectories in a complete 10$^{17}$~eV air shower.
   }
   \end{figure}

The shower evolution is thus taken into account in a continuous and consistent way by random creation of particles with probabilities according to eq.\ (\ref{eqn:inject}). Fig.\ \ref{fig:trajectrace} illustrates the shower evolution through the particle trajectories that are followed during the simulation of our $10^{17}$~eV vertical air shower.

\subsection{Integrated shower results}

\begin{figure}
   \psfrag{EmaxmuVpm}[c][t]{Max$\left(\left|\vec{E}(t)\right|\right)$~[$\mu$V m$^{-1}$]}
   \includegraphics[height=8.6cm,angle=270]{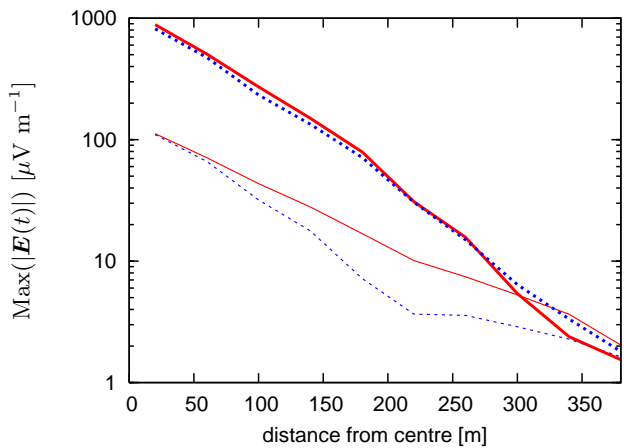}
   \caption{
   \label{fig:maxampasymmetryintegrated}
   Effects introduced in the radial dependence of the 40--160~MHz rectangle-filtered pulse amplitude by the integration over the shower evolution. Thick lines: integrated $10^{17}$~eV vertical shower with broken power-law particle energy distribution, statistical track length distribution with $\lambda=40$~g~cm$^{-2}$ and 70$^{\circ}$ inclined 0.5~Gauss magnetic field, thin lines: only shower maximum in same scenario. Solid: to the north, dashed: to the west.
   }
   \end{figure}

\begin{figure}
   \psfrag{Eew0muVpm}[c][t]{$E_{\mathrm{EW}}(t)$~[$\mu$V m$^{-1}$]}
   \includegraphics[height=8.6cm,angle=270]{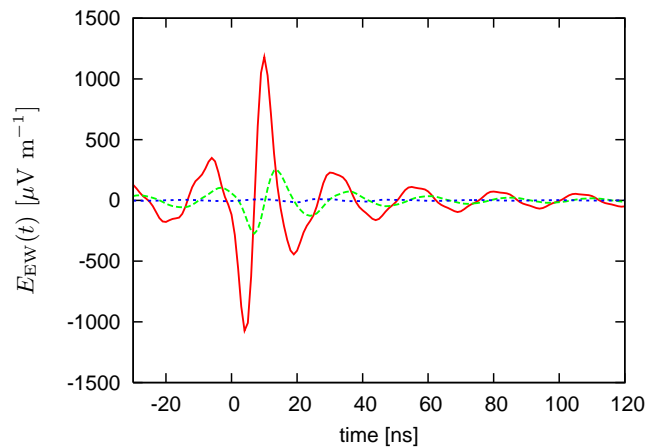}
   \caption{
   \label{fig:pulses0m100m260m}
   Pulses in the east-west polarisation component after 40--160~MHz rectangle-filtering for a shower as described in Fig.\ \ref{fig:maxampasymmetryintegrated}. Solid: in the shower centre, long dashed: 100~m to north of centre, short dashed: 260~m to north of centre.
   }
   \end{figure}

As in the theoretical calculations, the integration over the shower evolution has two main effects, visible in Fig.\ \ref{fig:maxampasymmetryintegrated}: First, the emission level is boosted significantly. This directly shows that the emission is not described sufficiently by just taking into account the shower maximum. The second major effect is a steepening of the radial emission pattern due to the amplification of coherence losses. It is mainly the centre region which receives significant additional radiation. The steepness of the radial dependence is also illustrated by the strong drop in the maximum amplitude of the filtered pulses when one goes to even moderate distances of 260~m as shown in Fig.\ \ref{fig:pulses0m100m260m}.\footnote{The integrated pulse shown in Fig.\ 18 of \protect\citet{HuegeFalcke2003a} was by mistake calculated for a 40--160~MHz rectangle filter. Using the intended 42.5--77.5~MHz rectangle filter, the amplitude drops from $\sim1500$~$\mu$V~m$^{-1}$ to $\sim1100$~$\mu$V~m$^{-1}$.}

\begin{figure*}
   \includegraphics[width=4.1cm,angle=270]{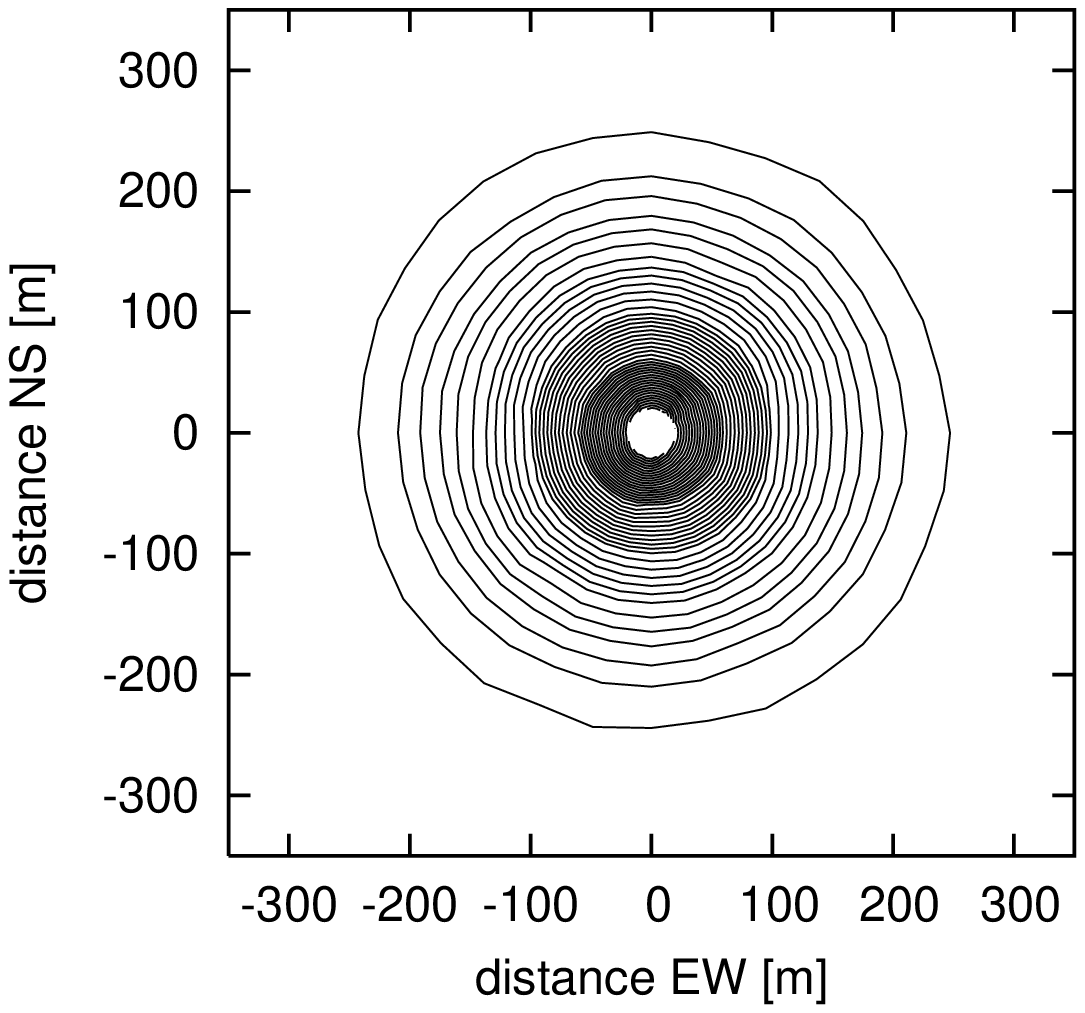}
   \includegraphics[width=4.1cm,angle=270]{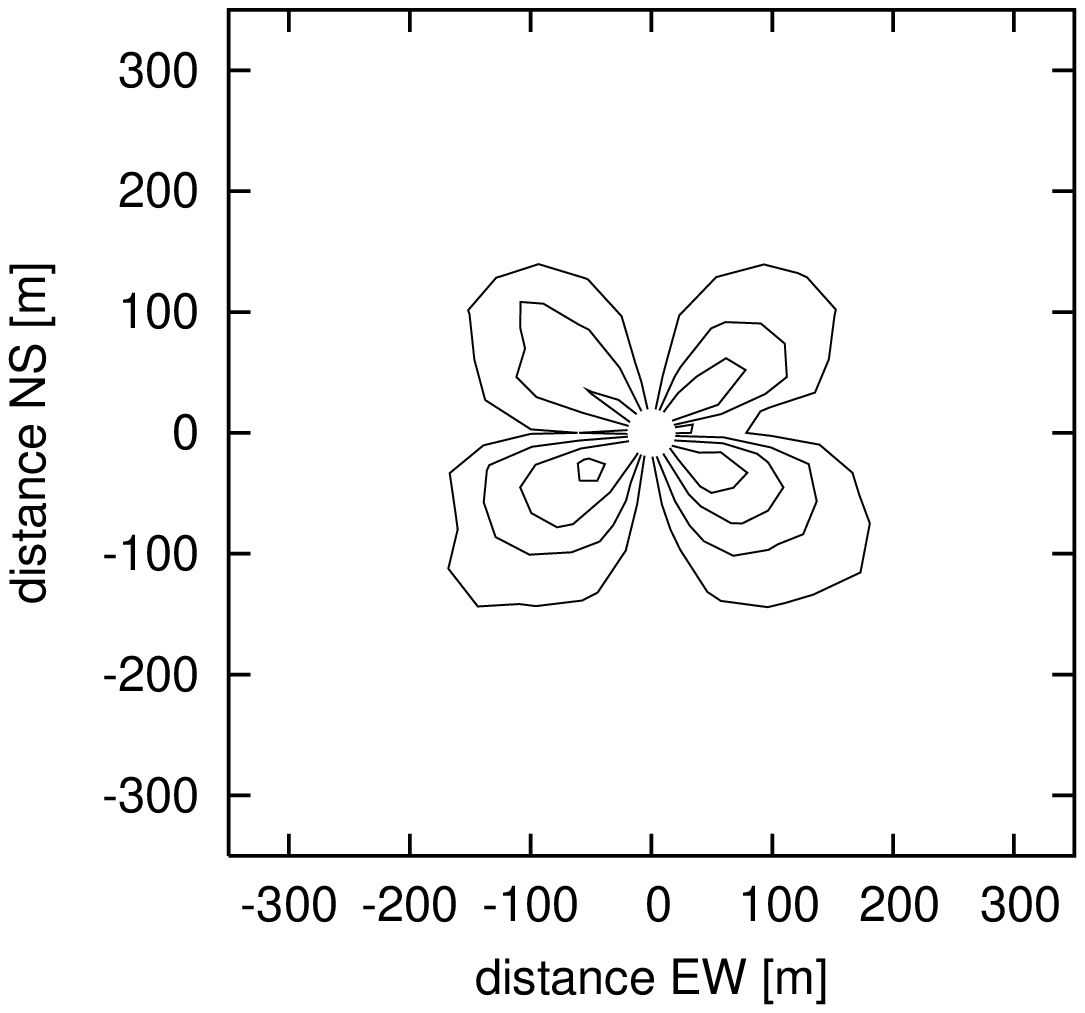}
   \includegraphics[width=4.1cm,angle=270]{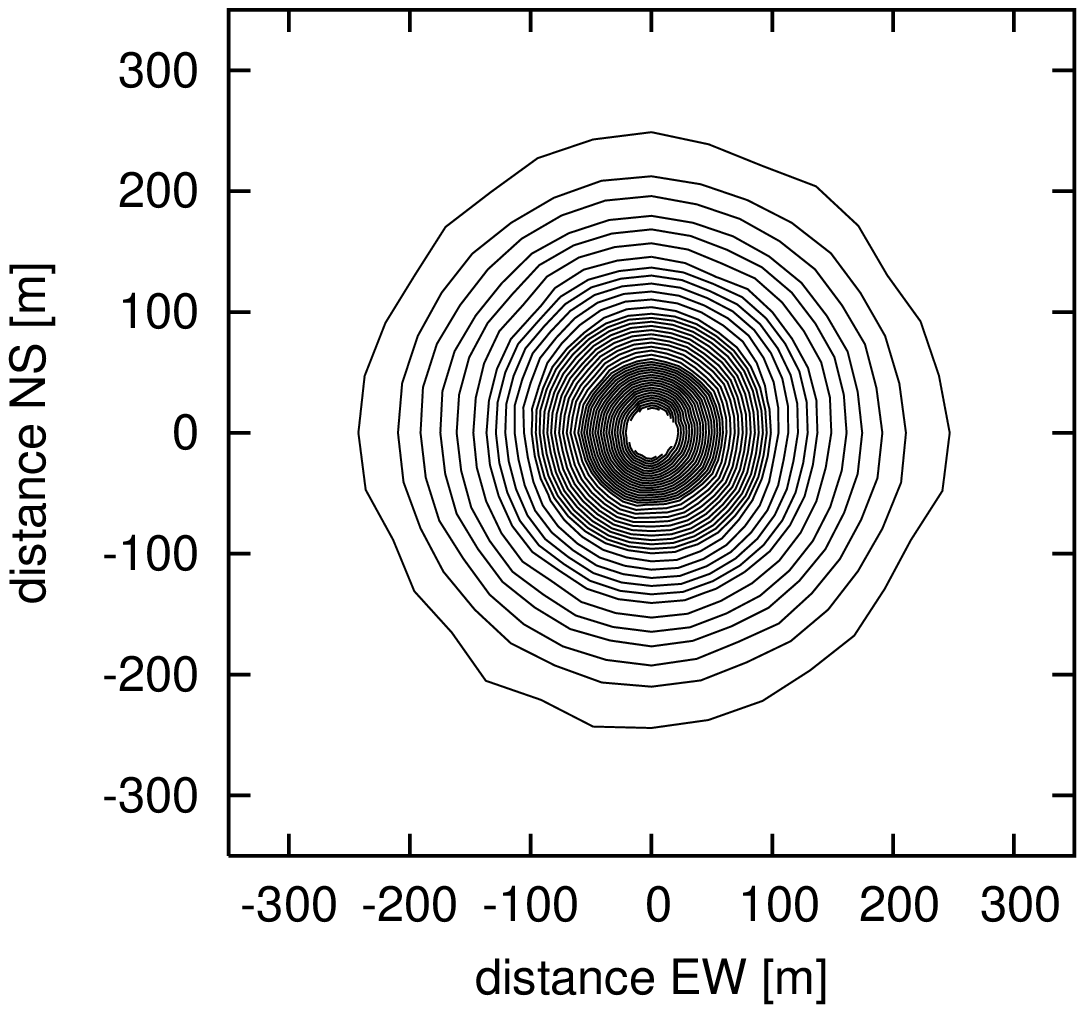}
   \includegraphics[width=4.1cm,angle=270]{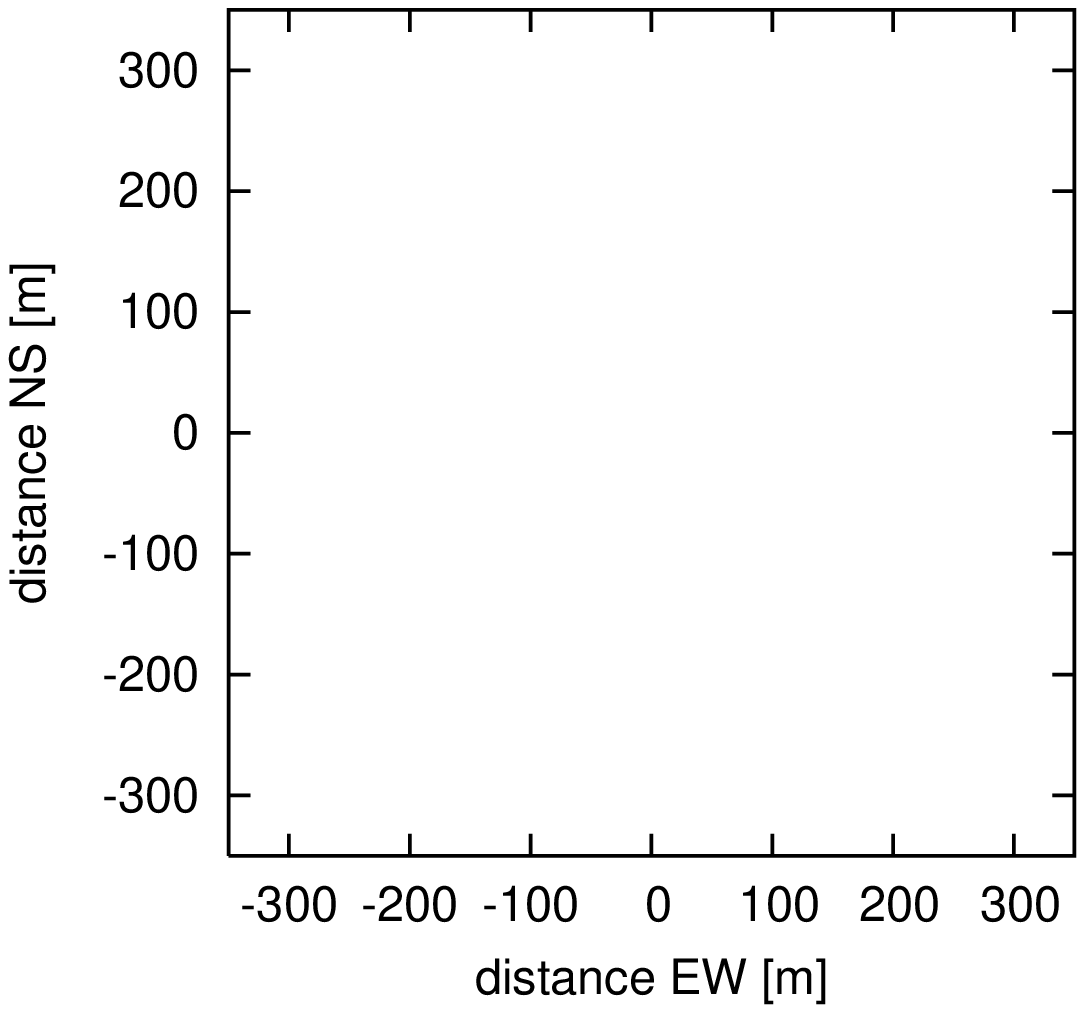}
   \caption{
   \label{fig:real05contours}
   Contour plots of the 40--160~MHz rectangle-filtered pulse amplitude for the same scenario as described in Fig.\ \ref{fig:maxampasymmetryintegrated}. Contour levels are 20~$\mu$V~m$^{-1}$ apart. From left to right: total electric field strength, north-south polarisation component, east-west polarisation component, vertical polarisation component.
   }
   \end{figure*}

Another important effect is the further fading away of sharp features and the asymmetries associated with the geomagnetic field in the emission pattern, as illustrated by Fig.\ \ref{fig:real05contours}. While some of these features were still quite prominent in the emission from a single shower slice, they more or less vanish completely as soon as the integration is performed. Similarly, the effects of individual changes to the particle distributions, such as the specific choice of an energy distribution, are therefore much weaker than they are in case of a single shower slice.

\begin{figure}
   \psfrag{Exyomega0muVpmpMHz}[c][t]{$E_{x/y}(\vec{R},\omega)$~[$\mu$V~m$^{-1}$~MHz$^{-1}$]}   
   \includegraphics[height=8.6cm,angle=270]{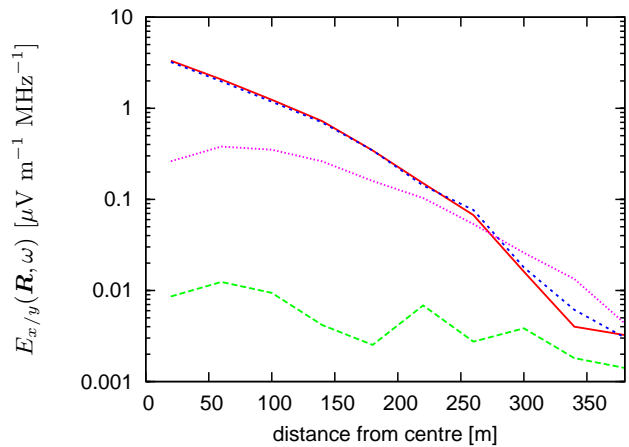}
   \caption{
   \label{fig:polarisationasymmetry}
   Radial dependence of $E(\vec{R},\omega)$ for different polarisation components at $\nu=55$~MHz for the same scenario as in Fig.\ \ref{fig:maxampasymmetryintegrated}. Solid: east-west polarisation to the north from centre, long-dashed: north-south polarisation to the north from centre, short-dashed: east-west polarisation to the north-west from centre, dotted: north-south polarisation to the north-west from centre.
   }
   \end{figure}

The polarisation characteristics of the integrated shower are very similar to those of an individual slice. Again, the emission is highly polarised in the east-west direction, except for some regions at medium distances of $\sim$200--400~m as illustrated in Fig.\ \ref{fig:real05contours} and, in a more quantitative way, Fig.\ \ref{fig:polarisationasymmetry}.

\begin{figure}
   \psfrag{EmaxvsRhonkg}[c][t]{Max$\left(\left|\vec{E}(t)\right|\right)$ versus $\rho_{\mathrm{NKG}}$}   
   \includegraphics[height=8.6cm,angle=270]{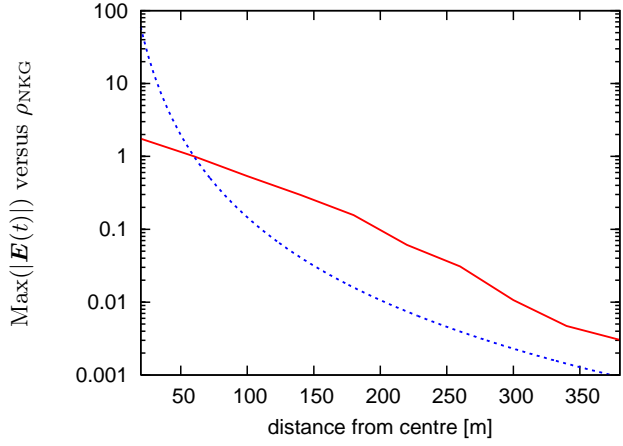}
   \caption{
   \label{fig:maxampvsnkg}
   Comparison of the filtered radio pulse amplitude (solid) and the particle density as given by the NKG-parametrisation (dashed) as a function of radial distance from the shower centre. Values were normalised to unity at $r=60$~m.
   }
   \end{figure}

Figure \ref{fig:maxampvsnkg} shows a direct comparison of the radial profiles of the particle density (as given by the NKG-parametrisation) and the filtered radio pulse amplitude. The particle density falls off much more steeply than the radio signal in the central $\sim$200--250 metres. Further out, the slope becomes comparable for radio amplitude and particle density.

\subsection{Comparison with theoretical calculations}

\begin{figure}
   \psfrag{Eomegaew0muVpmpMHz}[c][t]{$E_{\mathrm{EW}}(\vec{R},2\pi\nu)$~[$\mu$V~m$^{-1}$~MHz$^{-1}$]}   
   \psfrag{nu0MHz}[c][t]{$\nu$~[MHz]}
   \includegraphics[height=8.6cm,angle=270]{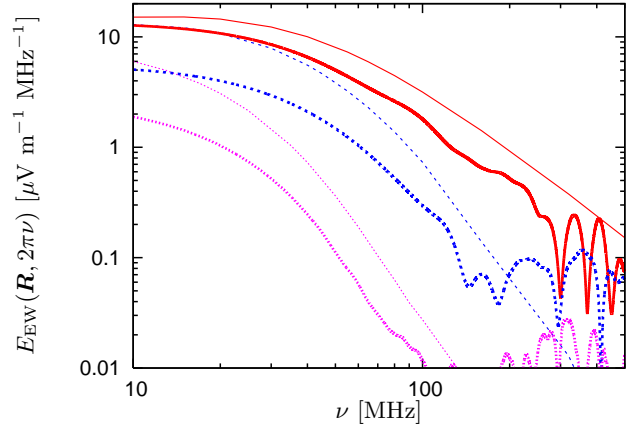}
   \caption{
   \label{fig:speccomp_fullshower_factor1.0}
   Spectra emitted by a complete $10^{17}$~eV vertical air shower with maximum at 4~km height, broken power-law particle energy distribution, statistically distributed particle trajectories with 40~g~cm$^{-2}$ mean path length and 70$^{\circ}$ inclined 0.5~Gauss magnetic field. Thin lines: analytic calculations of \protect\citet{HuegeFalcke2003a}, thick lines: these MC simulations. Solid: shower centre, dashed: 100~m, dotted: 250~m to north from shower centre.
   }
   \end{figure}

\begin{figure}
   \psfrag{Eomegaew0muVpmpMHz}[c][t]{$E_{\mathrm{EW}}(\vec{R},2\pi\nu)$~[$\mu$V~m$^{-1}$~MHz$^{-1}$]}   
   \psfrag{nu0MHz}[c][t]{$\nu$~[MHz]}
   \includegraphics[height=8.6cm,angle=270]{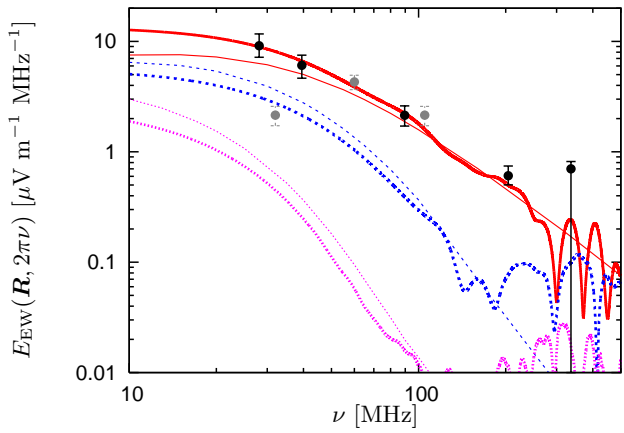}
   \caption{
   \label{fig:speccomp_fullshower_factor2.0}
   Same as Fig.\ \ref{fig:speccomp_fullshower_factor1.0} but scaling down the analytic results by a factor of two. Data from \protect\citet{Prah1971} (gray) and \protect\citet{Spencer1969} (black) were rescaled to be consistent with the \protect\citet{Allan1971} data at $\nu=55$~MHz.
   }
   \end{figure}

We now compare the results of our MC simulations of a fully integrated $10^{17}$~eV vertical air shower with the theoretical calculations performed in \citet{HuegeFalcke2003a}. Fig.\ \ref{fig:speccomp_fullshower_factor1.0} shows the spectral dependence of the emission in direct comparison. The MC results again produce somewhat lower levels of radiation. Scaling down the theoretical results by the systematic factor of two introduced in section \ref{sec:slicecomparison}, the agreement is much better, as shown in Fig.\ \ref{fig:speccomp_fullshower_factor2.0}. Considering the huge differences in the two calculations, the agreement is quite remarkable.

In \citet{HuegeFalcke2003a} we compared the theoretical results with the available historical data. As discussed there in detail, the absolute values of the historical experimental data are very uncertain and largely discrepant between the different groups. Additional uncertainty arises from ambiguities in the exact definition of the historical values denoted as $\epsilon_{\nu}$ and their conversion to the theoretical quantity $\left|\vec{E}(\vec{R},\omega)\right|$, which we performed via the relation
\begin{equation} \label{eqn:econversion}
\epsilon_{\nu}=\sqrt{\frac{128}{\pi}}\left|\vec{E}(\vec{R},\omega)\right| \approx 6.4 \left|\vec{E}(\vec{R},\omega)\right|.
\end{equation}
To deal with the discrepant sets of data we decided to take the well documented \citet{Allan1971} data as our reference and rescale the spectral data of \citet{Prah1971} and \citet{Spencer1969} to be consistent with the Allan data at $\nu=55$~MHz. In this work, we use the identically rescaled data for comparison with our new MC spectra. While the absolute values of the spectral data are therefore somewhat arbitrary, they at least allow an evaluation of the qualitative spectral dependence. The spectral data are over-plotted in Fig. \ref{fig:speccomp_fullshower_factor2.0}. The agreement with the spectral dependence in the centre region of the shower is very good.

\begin{figure}
   \psfrag{Eomegaew0muVpmpMHz}[c][t]{$E_{\mathrm{EW}}(\vec{R},2\pi\nu)$~[$\mu$V~m$^{-1}$~MHz$^{-1}$]}   
   \includegraphics[height=8.6cm,angle=270]{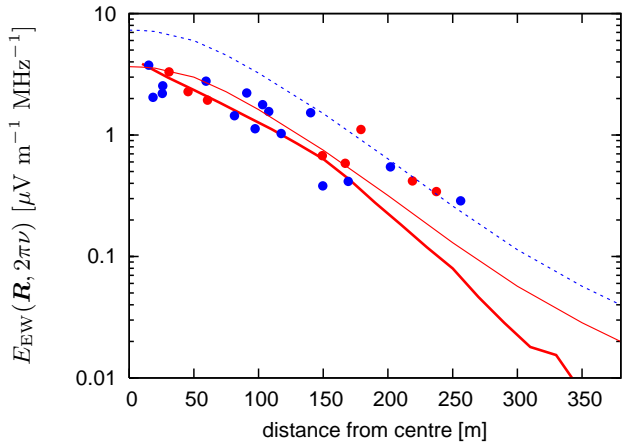}
   \caption{
   \label{fig:radialcomp_fullshower}
   Radial dependence of the emission at $\nu=55$~MHz for the same scenario as in Fig.\ \ref{fig:speccomp_fullshower_factor1.0}. Dashed: analytic calculations of \protect\citet{HuegeFalcke2003a}. Solid: analytic calculations of \protect\citet{HuegeFalcke2003a} scaled down by a factor of two (thin line) in comparison with these MC simulations (thick line). Data from \protect\citet{Allan1971}. Distance is to north from shower centre.
   }
   \end{figure}

In Fig.\ \ref{fig:radialcomp_fullshower} we compare the radial dependence of the emission as calculated by our MC code with the earlier theoretical results and the \citet{Allan1971} data. The scaled-down theoretical results again show good agreement with the MC results in the centre region. Overall, the MC predicts a somewhat steeper decrease of the emission levels to higher distances. It was, however, expected that the theoretical calculations overestimate the coherence and thus emission levels at high distances. The absolute level of the emission also agrees with the Allan data within the uncertainties.

\section{Discussion}

With the design and implementation of our MC simulation, we have taken the logical next step in the development of our geosynchrotron radiation model for radio emission from cosmic ray air showers. The MC technique provides an independent cross-check on our earlier theoretical works and allows us to model the emission from the air shower with much higher precision based on a much more realistic air shower model.

To make these simulations feasible on standard ``off-the-shelf'' computer hardware, we conceived and implemented a number of intelligent concepts. In particular, the smart sampling of the particle trajectories in conjunction with the automatic inactivation of ground-bins and the adaptive collection of the time-series data provide the necessary cut-down on computation time compared to a simple ``brute-force'' approach. At the same time, high precision is retained in the results as demonstrated by the detailed consistency checks presented in section \ref{sec:consistencychecks}. The calculation of a typical vertical $10^{17}$ eV air shower with realistic shower properties and $25,000,000$ particles combined with automatic ground-bin inactivation at very high precision takes about 400 seconds per ground-bin on a standard PC of the 1.6~GHz class. A very detailed calculation with 800 ground-bins (25 radial, 32 azimuthal), parallelised on 4 PCs, thus takes about one day. 

A major new result of our MC simulations is the profound influence of edge effects arising from the finite lengths of the particle trajectories. These could not be taken into account in the theoretical calculations. In combination with the dependence of the individual particle emission on the magnetic field strength, the edge effects associated with the statistical distribution of track lengths wash out the asymmetries originally introduced into the total field strength emission pattern by the geomagnetic field to a high degree. Once the integration over the shower evolution as a whole is carried out, the asymmetry is gone completely. The absence of asymmetries in the total field strength emission pattern from an integrated shower is somewhat unfortunate, as a prominent asymmetry would have allowed an easy, yet unambiguous test of the geomagnetic emission mechanism via the statistics of radio pulse total field strengths alone.

The decomposition of the electric field into north-south, east-west and vertical polarisation components, however, shows that the emission is indeed highly polarised in the direction perpendicular to the shower axis and the magnetic field direction, as predicted by our theoretical calculations. Polarisation-dependent radio measurements such as the ones carried out by LOPES could therefore still unambiguously establish the geomagnetic emission mechanism.

The MC results show good consistency with the theoretical spectra and radial dependences --- apart from a systematic factor of two in emission strength which is plausible considering the implicit assumption of symmetric trajectories in the analytic calculations. Such good agreement between the theoretical and MC calculations is remarkable considering the inherent differences in the two approaches. In particular, the integration over the shower evolution as a whole is carried out in a much more sophisticated way in the MC simulations as compared to the analytical works. As mentioned above, the polarisation characteristics of the emission are exactly as inferred in the theoretical calculations.

The spectra and radial dependences predicted by our MC code also agree well with the historical data of \citet{Allan1971} and the data of \citet{Prah1971} and \citet{Spencer1969} scaled to the absolute level of the Allan data. The necessary rescaling, however, demonstrates that the historical data themselves are largely discrepant with absolute values reaching up to an order of magnitude lower than the Allan data (see \citet{HuegeFalcke2003a} for a detailed discussion). It is therefore still imperative to gather independent data with good absolute calibration with new experiments such as LOPES \citep{HornefferFalckeHaungs2003}. For air showers of energies around 10$^{17}$~eV this should be well feasible as the predicted absolute levels of emission $|\vec{E}(\vec{R},\omega)|$ around a few $\mu$V~m$^{-1}$~MHz$^{-1}$ at 55~MHz are well above the galactic noise limit of $\sim$0.4/0.15/0.05~$\mu$V~m$^{-1}$~MHz$^{-1}$ for a 3$\sigma$ detection with 1/10/100 LOPES antenna(s) \citep{HuegeFalcke2003a}. In areas with high radio-frequency interference levels such as the site of the KASCADE array, the noise levels are a factor of a few higher.

Further major improvement of our model will be reached by the interfacing of our MC simulation to the air shower simulation code CORSIKA \citep{HeckKnappCapdevielle1998}. The solid particle physics foundation of CORSIKA will resolve the uncertainties related to the somewhat crude shower parametrisations used so far. (Detailed MC calculations of the RICE group \citep{RazzaqueSeunarineBesson2002,RazzaqueSeunarineChambers2004}, e.g., reveiled that the compactness of the particle distributions for electromagnetic showers in ice --- and thus the coherence of the radio emission at high frequencies --- was originally underestimated in analytical parametrisations.)

\section{Conclusions}

We have successfully advanced our modelling of radio emission from cosmic ray air showers with elaborate Monte Carlo simulations in the time-domain. Our MC code takes into account the important air shower characteristics such as lateral and longitudinal particle distributions, particle energy and track length distributions, a realistic magnetic field geometry and the evolution of the air shower as a whole. The calculation retains the full polarisation information and does not employ any far-field approximations.

We predict emission patterns, radial and spectral dependences for an exemplary $10^{17}$~eV vertical air shower and find good agreement with our earlier theoretical works and the historical data available.

A major result that could not be obtained by analytic calculations alone is that asymmetries introduced into the total field strength emission pattern by the magnetic field direction are washed out completely in the radiation from an integrated air shower. Statistics of total field strengths alone can therefore not establish the geomagnetic emission mechanism. The clear polarisation dependence on the magnetic field direction, on the other hand, allows a direct test of the geomagnetic emission mechanism through polarisation-sensitive experiments such as LOPES.

After having documented the implementation details and having demonstrated the correctness and robustness of our MC simulations, our code is now in a stage where we can explore the dependence of the radio emission on a number of parameters such as shower axis direction, primary particle energy, depth of shower maximum and the like. Consequently, this will be our next step.

Once these dependences are established, measurements of radio emission from cosmic ray air showers can be related directly to the underlying characteristics of the observed air showers. Due to the regularity and robustness of the modelled emission patterns, even a sparse sampling of the radiation pattern with a limited number of antennas would probably suffice for such an analysis.

Furthermore, our code provides a solid basis for the inclusion of additional effects such as Askaryan-type \citep{Askaryan1962a,Askaryan1965} \v{C}erenkov radiation and an interfacing to the MC air shower simulation code CORSIKA.

\begin{acknowledgements}
We would like to thank Elmar K\"ording and Andreas Horneffer for numerous useful discussions. Additional thanks go to John Ralston, Ralf Engel and Klaus Werner for helpful comments. We are also grateful for the useful comments of the anonymous referee. LOPES is supported by the German Federal Ministry of Education and Research under grant No.\ 05 CS1ERA/1 (Verbundforschung Astroteilchenphysik).
\end{acknowledgements}

\section*{Appendix: Trajectory}

Consider first a simple magnetic field geometry
\begin{equation}
\vec{\tilde{B}} = B\,\left( \begin{array}{c} 0\\ 0\\ 1 \end{array} \right).
\end{equation}
The (unperturbed) trajectory of a charged particle in a homogeneous magnetic field is a helix. Aligning the helix along the $z$-axis, it can be written as
\begin{equation} \label{eqn:simpletrajectory}
\vec{\tilde{r}}(t)= \left( \begin{array}{c}
- R_{B} \cos \left[\omega_{B} (t-t_{0})\right] \\
+ R_{B} \sin \left[\omega_{B} (t-t_{0})\right] \\
  v \cos{\alpha}\ (t-t_{0})
\end{array} \right),
\end{equation}
where
\begin{equation}
R_{B} = \frac{v \sin{\alpha}}{\omega_{B}}
\end{equation}
denotes the radius of the projected circular motion and
\begin{equation}
\omega_{B} = \frac{q e B}{\gamma m_{e} c}
\end{equation}
is the gyration frequency of the particle with charge $q$ times the elementary charge unit $e$ and velocity
\begin{equation}
v = \beta\ c = \sqrt{1-\frac{1}{\gamma^{2}}}\ c.
\end{equation}
The pitch-angle $\alpha$ is given by the (constant) angle between the direction of the particle velocity vector and the magnetic field vector.

To derive the general form of this trajectory, we first rotate the coordinate system such that the $B$-field points in the desired direction. Afterwards, we adjust the phase $t_{0}$ such that the particle's initial velocity  has the desired direction as specified by the initial velocity vector $\vec{V}$. In the last step we shift the trajectory so that at $t=0$ it coincides with the desired starting position $\vec{R}$.

We want to transform the simple geometry field $\vec{\tilde{B}}$ to the desired geometry
\begin{equation}
\vec{B} = B\,\left( \begin{array}{c}
\cos{\theta} \sin{\varphi} \\
\sin{\theta} \sin{\varphi} \\
\sin{\theta}
\end{array} \right),
\end{equation}
where $\varphi \in [0,2\pi[$ and $\theta \in [0,\pi]$ are the azimuth and zenith angles known from spherical coordinates. This transformation is achieved by applying a rotation matrix
\begin{eqnarray}
\mathbf{D} &=& \left( \begin{array}{ccc}
\cos{\varphi}	&	-\sin{\varphi}	&	0	\\
\sin{\varphi}	&	\cos{\varphi}	&	0	\\
0				&	0				&	1	\\
\end{array} \right)
\cdot
\left( \begin{array}{ccc}
\cos{\theta}	&	0				&	\sin{\theta}	\\
0				&	1				&	0				\\
-\sin{\theta}	&	0				&	\cos{\theta}	\\
\end{array} \right)	\nonumber \\
&=& \left( \begin{array}{ccc}
\cos{\theta} \cos{\varphi}	&	-\sin{\varphi}			&	\sin{\theta} \cos{\varphi}	\\
\cos{\theta} \sin{\varphi}	&	\cos{\varphi}			&	\sin{\theta} \sin{\varphi}	\\
-\sin{\theta}				&	0						&	\cos{\theta}
\end{array} \right),
\end{eqnarray}
so
\begin{equation}
\vec{B} = \mathbf{D} \vec{\tilde{B}}
\end{equation}
and, inversely,
\begin{equation}
\vec{\tilde{B}} = \mathbf{D}^{-1} \vec{B}.
\end{equation}
Applying the same rotation to the trajectory (\ref{eqn:simpletrajectory}) yields
\begin{equation} \label{eqn:trajecr}
\vec{r}(t) = \mathbf{D}\ \vec{\tilde{r}}(t).
\end{equation}
To infer the phase $t_{0}$ corresponding to a given initial velocity $\vec{V}$, we rotate back $\vec{V}$ to the simple geometry,
\begin{equation}
\vec{\tilde{V}} = \mathbf{D}^{-1} \vec{V}.
\end{equation}
The $x$- and $y$-component of $\vec{\tilde{V}}$ then directly determine $t_{0}$ through the relation
\begin{equation}
\omega_{B} t_{0} = \mp \arctan \left(\pm \frac{\tilde{V}_{x}}{\tilde{V}_{y}}\right),
\end{equation}
with the upper sign for $q>0$ and the lower for $q<0$ and where one has to take into account the correct quadrant for the $\arctan$ operation. In terms of the components of $\vec{V}$ this yields
\begin{equation} \label{eqn:traject0}
t_{0} = \mp \arctan\left(\frac{\pm \left[(V_{x} \cos{\varphi} + V_{y} \sin{\varphi}) \cos{\theta}-V_{z} \sin{\theta}\right]}
{V_{y} \cos{\varphi} - V_{x} \sin{\varphi}}˜\right) / \omega_{B}.
\end{equation}
The last operation that has to be employed is a translation
\begin{equation} \label{eqn:trajecrabs}
\vec{r}_{\mathrm{abs}}(t) = \vec{R}_{0} + \vec{r}(t)
\end{equation}
of the trajectory such that
\begin{equation}
\vec{r}_{\mathrm{abs}}(t=0) = \vec{R}_{0} + \vec{r}(t=0) = \vec{R},
\end{equation}
which yields
\begin{eqnarray} \label{eqn:trajecr0}
\vec{R}_{0} &=& \vec{R} + v \cos{\alpha}\ t_{0} \left(
\begin{array}{c}
\sin{\theta} \cos{\varphi} \\
\sin{\theta} \sin{\varphi} \\
\cos{\theta}
\end{array} \right)	\nonumber \\
&+&
R_{B} \left( \begin{array}{c}
\cos{\theta} \cos{\varphi} \cos[\omega_{B} t_{0}] - \sin{\varphi} \sin[\omega_{B} t_{0}]	\\
\cos{\theta} \sin{\varphi} \cos[\omega_{B} t_{0}] + \cos{\varphi} \sin[\omega_{B} t_{0}]	\\
-\sin{\theta} \cos[\omega_{B} t_{0}]
\end{array} \right).
\end{eqnarray}
The resulting trajectory $\vec{r}_{\mathrm{abs}}(t)$ is thus fully defined for a given set of parameters $\vec{R}$ and $\vec{V}$. The time-dependence of particle velocity and acceleration are then easily derived as the time-derivatives of $\vec{r}_{\mathrm{abs}}(t)$.
%


\end{document}